\documentclass[longauth]{aa}

\usepackage[switch]{lineno}
\nolinenumbers
\usepackage{graphicx}
\usepackage{arydshln}
\usepackage{subcaption}
\usepackage{hyperref}
\hypersetup{colorlinks=true,linkcolor=blue,filecolor=blue,urlcolor=blue,citecolor=blue}
\usepackage{appendix}
\usepackage{comment}
\usepackage{anyfontsize}
\usepackage[normalem]{ulem}
\usepackage{txfonts}
\makeatletter
\renewcommand*\aa@pageof{, page \thepage{} of \pageref*{LastPage}}
\makeatother
\usepackage{multirow}
\usepackage{multicol}
\usepackage{orcidlink}

\begin{document}

\title{Insights into the broad-band emission of the TeV blazar Mrk\,501 during the first X-ray polarization measurements}

\titlerunning{Radio to TeV observations of Mrk~5011 during the \textit{IXPE} campaigns}
%
\author{
\small
S.~Abe\inst{1} \and
J.~Abhir\inst{2}\orcidlink{0000-0001-8215-4377} \and
V.~A.~Acciari\inst{3} \orcidlink{0000-0001-8307-2007} \and
A.~Aguasca-Cabot\inst{4} \orcidlink{0000-0001-8816-4920} \and
I.~Agudo\inst{5} \orcidlink{0000-0002-3777-6182} \and
T.~Aniello\inst{6} \orcidlink{} \and
S.~Ansoldi\inst{7,44} \orcidlink{0000-0002-5613-7693} \orcidlink{} \and
L.~A.~Antonelli\inst{6}\orcidlink{0000-0002-5037-9034} \orcidlink{} \and
A.~Arbet Engels\inst{8}{$^\star$}\orcidlink{0000-0001-9076-9582} \orcidlink{} \and
C.~Arcaro\inst{9} \orcidlink{0000-0002-1998-9707} \orcidlink{}
% ...
}
\institute { Japanese MAGIC Group: Institute for Cosmic Ray Research (ICRR), The University of Tokyo, Kashiwa, 277-8582 Chiba, Japan
\and ETH Z\"urich, CH-8093 Z\"urich, Switzerland
\and Instituto de Astrof\'isica de Canarias and Dpto. de  Astrof\'isica, Universidad de La Laguna, E-38200, La Laguna, Tenerife, Spain
\and Universitat de Barcelona, ICCUB, IEEC-UB, E-08028 Barcelona, Spain
\and Instituto de Astrof\'isica de Andaluc\'ia-CSIC, Glorieta de la Astronom\'ia s/n, 18008, Granada, Spain
\and National Institute for Astrophysics (INAF), I-00136 Rome, Italy
\and Universit\`a di Udine and INFN Trieste, I-33100 Udine, Italy
\and Max-Planck-Institut f\"ur Physik, D-80805 M\"unchen, Germany
\and Universit\`a di Padova and INFN, I-35131 Padova, Italy
\and Croatian MAGIC Group: University of Zagreb, Faculty of Electrical Engineering and Computing (FER), 10000 Zagreb, Croatia
\and IPARCOS Institute and EMFTEL Department, Universidad Complutense de Madrid, E-28040 Madrid, Spain
\and Centro Brasileiro de Pesquisas F\'isicas (CBPF), 22290-180 URCA, Rio de Janeiro (RJ), Brazil
\and University of Lodz, Faculty of Physics and Applied Informatics, Department of Astrophysics, 90-236 Lodz, Poland
\and Centro de Investigaciones Energ\'eticas, Medioambientales y Tecnol\'ogicas, E-28040 Madrid, Spain
\and Institut de F\'isica d'Altes Energies (IFAE), The Barcelona Institute of Science and Technology (BIST), E-08193 Bellaterra (Barcelona), Spain
\and Departament de F\'isica, and CERES-IEEC, Universitat Aut\`onoma de Barcelona, E-08193 Bellaterra, Spain
\and Universit\`a di Pisa and INFN Pisa, I-56126 Pisa, Italy
\and Department for Physics and Technology, University of Bergen, Norway
\and INFN MAGIC Group: INFN Sezione di Torino and Universit\`a degli Studi di Torino, I-10125 Torino, Italy
\and INFN MAGIC Group: INFN Sezione di Catania and Dipartimento di Fisica e Astronomia, University of Catania, I-95123 Catania, Italy
\and INFN MAGIC Group: INFN Sezione di Bari and Dipartimento Interateneo di Fisica dell'Universit\`a e del Politecnico di Bari, I-70125 Bari, Italy
\and Croatian MAGIC Group: University of Rijeka, Faculty of Physics, 51000 Rijeka, Croatia
\and Technische Universit\"at Dortmund, D-44221 Dortmund, Germany
\and University of Geneva, Chemin d'Ecogia 16, CH-1290 Versoix, Switzerland
\and Japanese MAGIC Group: Physics Program, Graduate School of Advanced Science and Engineering, Hiroshima University, 739-8526 Hiroshima, Japan
\and Deutsches Elektronen-Synchrotron (DESY), D-15738 Zeuthen, Germany
\and Armenian MAGIC Group: ICRANet-Armenia, 0019 Yerevan, Armenia
\and Croatian MAGIC Group: University of Split, Faculty of Electrical Engineering, Mechanical Engineering and Naval Architecture (FESB), 21000 Split, Croatia
\and Universit\"at W\"urzburg, D-97074 W\"urzburg, Germany
\and Croatian MAGIC Group: Josip Juraj Strossmayer University of Osijek, Department of Physics, 31000 Osijek, Croatia
\and Finnish MAGIC Group: Finnish Centre for Astronomy with ESO, University of Turku, FI-20014 Turku, Finland
\and Japanese MAGIC Group: Department of Physics, Tokai University, Hiratsuka, 259-1292 Kanagawa, Japan
\and Universit\`a di Siena and INFN Pisa, I-53100 Siena, Italy
\and Saha Institute of Nuclear Physics, A CI of Homi Bhabha National Institute, Kolkata 700064, West Bengal, India
\and Inst. for Nucl. Research and Nucl. Energy, Bulgarian Academy of Sciences, BG-1784 Sofia, Bulgaria
\and Japanese MAGIC Group: Department of Physics, Yamagata University, Yamagata 990-8560, Japan
\and Finnish MAGIC Group: Space Physics and Astronomy Research Unit, University of Oulu, FI-90014 Oulu, Finland
\vfill\null
\and Japanese MAGIC Group: Chiba University, ICEHAP, 263-8522 Chiba, Japan
\and Japanese MAGIC Group: Institute for Space-Earth Environmental Research and Kobayashi-Maskawa Institute for the Origin of Particles and the Universe, Nagoya University, 464-6801 Nagoya, Japan
\and Japanese MAGIC Group: Department of Physics, Kyoto University, 606-8502 Kyoto, Japan
\and INFN MAGIC Group: INFN Sezione di Perugia, I-06123 Perugia, Italy
\and INFN MAGIC Group: INFN Roma Tor Vergata, I-00133 Roma, Italy
\and Japanese MAGIC Group: Department of Physics, Konan University, Kobe, Hyogo 658-8501, Japan
\and also at International Center for Relativistic Astrophysics (ICRA), Rome, Italy
% ...
}

\date{Received XX XX, 2023; accepted XX XX, 2023}

\abstract
   {}
   {We present the first multi-wavelength study of Mrk 501 including very-high-energy (VHE) $\gamma$-ray observations simultaneous to X-ray polarization measurements from the Imaging X-ray Polarimetry Explorer (\textit{IXPE}). }
   {We use radio-to-VHE data from a multi-wavelength campaign organized between 2022-03-01 and 2022-07-19 (MJD~59639 to MJD~59779). The observations were performed by MAGIC, \textit{Fermi}-LAT, \textit{NuSTAR}, \textit{Swift} (XRT and UVOT), and several instruments covering the optical and radio bands to complement the \textit{IXPE} pointings. We characterize the dynamics of the broad-band emission around the X-ray polarization measurements through its multi-band fractional variability and correlations, and compare changes observed in the polarization degree to changes seen in the broad-band emission using a multi-zone leptonic scenario.}
   {During the \textit{IXPE} pointings, the VHE state is close to the average behavior with a 0.2-1 TeV flux of 20\% - 50\% the emission of the Crab Nebula. Additionally, it shows low variability and a hint of correlation between VHE and X-rays. Despite the average VHE activity, an extreme X-ray behavior is measured for the first two \textit{IXPE} pointings in March 2022 (MJD 59646 to 59648 and MJD 59665 to 59667) with a synchrotron peak frequency $>$\,1 keV. For the third \textit{IXPE} pointing in July 2022 (MJD 59769 to 59772), the synchrotron peak shifts towards lower energies and the optical/X-ray polarization degrees drop. All three \textit{IXPE} epochs show an atypically low Compton dominance in the $\gamma$-rays. The X-ray polarization is systematically higher than at lower energies, suggesting an energy-stratification of the jet. While during the \textit{IXPE} epochs the polarization angle in the X-ray, optical and radio bands align well, we find a clear discrepancy in the optical and radio polarization angles in the middle of the campaign. Such results further strengthen the hypothesis of an energy-stratified jet. We model the broad-band spectra simultaneous to the \textit{IXPE} pointings assuming a compact zone dominating in the X-rays and VHE, and an extended zone stretching further downstream the jet dominating the emission at lower energies. \textit{NuSTAR} data allow us to precisely constrain the synchrotron peak and therefore the underlying electron distribution. The change between the different states observed in the three IXPE pointings can be explained by a change of magnetization and/or emission region size, which directly connects the shift of the synchrotron peak to lower energies with the drop in polarization degree. }
   {}


   \maketitle

\section{Introduction} \label{sec:intro}
Blazars are among the brightest objects in the $\gamma$-ray sky and are known to emit radiation over a broad range of wavebands from radio up to the very-high-energy (VHE; $>0.1$\,TeV) regime. They are jetted active galactic nuclei AGNs with a small angle between our line of sight and the jet axis. 

Even though blazars have been observed for decades across the entire electromagnetic spectrum, their main acceleration and emission mechanisms are still debated. Polarization measurements in the different wavebands could distinguish between different emission mechanisms \citep{2013ApJ...774...18Z} or allow properties such as the underlying acceleration mechanisms or the magnetic field configurations to be probed \citep{2018MNRAS.480.2872T,2021Galax...9...37T}. Up to recently, polarization measurements of blazars have only been possible in the radio to optical bands. However, these bands are known for originating from more extended or multiple regions in the jet compared to the high-energy emission or are contaminated by the host galaxy \citep[see e.g.,][]{Mrk501_MAGIC_2014}. In December 2021, the Imaging X-ray Polarimetry Explorer (\textit{IXPE}) \citep{2022HEAD...1930101W} was launched with the goal of measuring linear polarization in the 2--8\,keV band for various sources. 

For the first time, the detection of X-ray polarization of the archetypal blazar Markarian\,501 \citep[Mrk\,501; z=0.034,][]{Ulrich_1975} was published by \citet{2022Natur.611..677L}. Mrk~501 is one of our closest and brightest blazars and therefore a prime object to explore new observational techniques. As usual for blazars, its spectral energy distribution (SED) depicts two peaks. The low-energy peak can be attributed to synchrotron radiation of relativistic electrons inside the magnetized plasma of the jet. It is used to classify blazars and Mrk\,501 usually is attributed to the class of high synchrotron peaked blazars (HSPs) with a synchrotron peak frequency $\nu_{s}>10^{15}$\,Hz \citep{2010ApJ...716...30A}. However, it has also shown extreme HSP (EHSP) behavior with $\nu_{s}\geq2.4\times10^{17}$\,Hz  ($\sim 1$\,keV) \citep{2001A&A...371..512C,2010ApJ...716...30A} during extended periods of time encompassing both flaring and non-flaring activity \citep[][]{Mrk501_MAGIC_2012, Mrk501_MAGIC_2013}. 
The origin of the high-energy peak is less clear. It could either originate from electron inverse-Compton scattering off low-energy photons \citep[leptonic scenarios; see, e.g.,][]{1992ApJ...397L...5M, 1996ASPC..110..436G, Tavecchio_Constraints} or could be induced by relativistic protons inside the jet sparking different cascades or hadronic synchrotron radiation \citep[hadronic scenarios; see, e.g.,][]{Mannheim93, Aharonian00, Anita_2001}.

For Mrk~501, the \textit{IXPE} energy range covers either the falling part, during its usual HSP states, or the peak, during EHSP behavior, of the synchrotron bump. \textit{IXPE} thus probes the emission from the most energetic electrons freshly accelerated in the jet.

The first X-ray polarization measurements of Mrk\,501 were conducted during two pointings from 2022-03-08 to 2022-03-10 and 2022-03-27 to 2022-03-29 (MJD 59646 to 59648 and MJD 59665 to 59667). In what follows, those two periods will be referred to as \textit{IXPE}-1 and \textit{IXPE}-2. They revealed a polarization degree in the 2--8\,keV band of $\sim10$\% that is a factor of two higher than in the optical regime \citep{2022Natur.611..677L}. Additionally, the polarization angle was found to be aligned with the radio jet. All those properties point towards shocks being the prime acceleration mechanism, as well as an energy stratified jet where the most energetic particles emit from a region which is magnetically more ordered and closer to the acceleration site \citep{2022Natur.611..677L, 2016MNRAS.463.3365A}. These energetic particles subsequently cool and stream away from the shock to populate broader regions. An additional \textit{IXPE} pointing was conducted from 2022-07-09 to 2022-07-12 (MJD 59769 to 59772) revealing a drop in polarization degree to $\sim7$\%, but with a stable polarization angle compared to the March measurement \citep{2023_IXPE_Mrk501}. This observing epoch is dubbed as \textit{IXPE}-3 throughout the rest of this paper.

In this work, we present the multi-wavelength (MWL) picture around these three X-ray polarization measurements, including, for the first time, also data up to the VHE regime taken by the \textit{Florian Goebel Major Atmospheric Gamma Imaging Cherenkov} (MAGIC) telescopes. To characterize the high-energy emission in detail, the campaign involved instruments such as the \textit{Neil Gehrels Swift Observatory (Swift)}, 
and the \textit{Nuclear Spectroscopic Telescope Array (NuSTAR)} in the X-rays, and the Large Area Telescope (LAT) on board the \textit{Fermi Gamma-ray Space Telescope} (\textit{Fermi}) in the high-energy $\gamma$-rays.
Both the \textit{IXPE} pointings as well as the whole time epoch from 2022-03-01 to 2022-07-19 (MJD 59639 to MJD 59779) were accompanied by extensive multi-wavelength coverage from radio to VHE, which is summarized in the following sections. 

This paper is structured as follows: In Section~\ref{sec:obs_analysis} we describe all instruments involved together with their dedicated data analysis methods. Section~\ref{sec:ixpe_mwl} provides a detailed characterization of the MWL emission during the three \textit{IXPE} observations, including the different flux states, spectral behaviors and polarization. The MWL characteristics of the full campaign are then summarized in Section~\ref{sec:longterm_MWL}.  Section~\ref{sec:modelling} presents a theoretical modeling of the three \textit{IXPE} time epochs using a two-zone leptonic model. Finally, Section~\ref{sec:discussion} summarizes and discusses the MWL results together with the obtained theoretical models. 

\section{Observations and analysis} \label{sec:obs_analysis}
A multitude of observations from VHE to radio were performed to investigate the full MWL behavior of Mrk\,501 around the first \textit{IXPE} observations. These were conducted in the framework of the extensive multi-instrument campaigns organized since 2008 for Mrk\,501 \citep{Mrk501_MAGIC_2008} and are stretching from 2022-03-01 to 2022-07-19 (MJD 59639 to MJD 59779) with the instruments and analyses summarized in this section.
\subsection{MAGIC}  \label{subsec:obs_analysis_magic}

MAGIC is a stereoscopic system of two imaging air cherenkov telescopes (IACTs) located at the Roque de los Muchachos Observatory, on the Canary island of La Palma at 2243\,m above sea level. MAGIC covers an energy range between tens of GeVs to tens of TeVs and has a sensitivity above 100 GeV (300 GeV) of about 2\% (about 1\%) of the Crab Nebula flux at low zenith angles ($<$~30$^\circ$) after 25~h of observations \citep[see Fig.~19 of][]{2016APh....72...76A}, which makes it well suited to perform blazar observations in the VHE range. 

Covering the time period around the first three \textit{IXPE} pointings, MAGIC observed Mrk\,501 for around 38\,hours from 2022-03-01 to 2022-07-19 (MJD 59639 to MJD 59779), which results in 26.5\, hours after data selection cuts based on atmospheric transmission. The data are analyzed using the MARS (MAGIC Analysis and Reconstruction Software) package \citep{zanin2013, 2016APh....72...76A}.

The VHE flux light curve (LC) is computed for the full energy range above 0.2\,TeV, as well as for a low-energy range of 0.2-1\,TeV and a high-energy range above 1\,TeV as shown in the top panel of Figure~\ref{fig:MWL_LC} using nightly bins. For the bins showing a significance of less than 2$\sigma$, upper limits (ULs) are computed according to the Rolke method \citep{Rolke_2005} with 95\% confidence. The spectral parameters around the three \textit{IXPE} pointings are determined using a forward folding method assuming a power-law model, which provides a good description of the data (see Table~\ref{tab:spec_par}). For the \textit{IXPE}-1 and \textit{IXPE}-2 spectra, the Tikhonov unfolding method is applied to compute the spectral data points \citep{2007unfold}, while for the \textit{IXPE}-3 spectrum again the forward folding method is used due to limited statistics. We checked the preference of a log parabola over a simple power-law model for each spectrum using a likelihood ratio test, but no significant ($>3\sigma$) preference is found. Extragalactic background light (EBL) absorption is corrected for using the model of~\citet{dominguez2011}.

\subsection{\textit{Fermi}-LAT}  \label{subsec:obs_analysis_fermi}

The LAT is a pair conversion instrument on board \textit{Fermi}, which operates in the high-energy regime and is sensitive to an energy range from 20\,MeV to beyond $300$\,GeV. It covers the entire sky every $3$\,hours \citep{2009ApJ...697.1071A,2012ApJS..203....4A}. 

To analyze the LAT data, we use the unbinned-likelihood tools from the \texttt{FERMITOOLS} software package\footnote{\url{https://fermi.gsfc.nasa.gov/ssc/data/analysis/}} (v2.0.8). We employ \texttt{P8R3\_SOURCE\_V3\_v1} as the instrument response function, \texttt{gll\_iem\_v07} \& \texttt{iso\_P8R3\_SOURCE\_V3\_v1} as the diffuse background\footnote{\url{http://fermi.gsfc.nasa.gov/ssc/data/access/lat/\\BackgroundModels.html}} and use the event class of 128, considering events that interacted in the front and back sections of the tracker (i.e. evtype = 3). The radius of the region of interest (ROI) is set to 10$^{\circ}$ and the maximum zenith to 100$^{\circ}$ because of the chosen energy range.  We chose a higher minimum energy than is conventionally chosen (0.3\,GeV instead of 0.1\,GeV) to ensure a better angular resolution (2 deg for the 68\% containment for photons above 0.3\,GeV compared to 5 deg for photons
above 0.1 GeV), which leads to an improvement in the signal-to-noise ratio since the analysis is less affected by diffuse backgrounds or potential (transient) non-accounted neighbouring sources. This is possible owing to the hard-spectrum of Mrk\,501 because the reduction of detected photons due to the higher threshold is relatively small. For the maximum energy, we chose a value of 500\,GeV that allows an overlap with the MAGIC energy range when reconstructing spectra.

To build a model, we use the fourth \textit{Fermi}-LAT source catalog \citep[4FGL-DL3;][]{2022ApJS..260...53A} using all sources within the ROI plus an annulus of 5$^\circ$. The obtained model is then fit to the data set in the time range  2022-03-01 to 2022-07-19 (MJD~59639 to MJD~59779). Subsequently, very weak components with a test statistic \citep[TS;][]{1996ApJ...461..396M} below 3 are removed from the model as well as those with an expected source count below 1. 
We then divide the data set in 20 bins 
of 7-day duration, and redo the fit in each bin varying only the normalization of the diffuse backgrounds, and that of sources within $<3^{\circ}$ and a detection with TS $>10$. Furthermore, we allow the spectral parameters of Mrk\,501 to vary assuming a power-law shape.
To evaluate the impact of very variable sources in our ROI, we conducted the same analysis freeing the normalization and spectral indices of all very-variable sources (variability index > 100). Both analysis agree within the statistical uncertainties.
In addition, spectral analyses are performed for 2 week intervals centered around each of the three \textit{IXPE} observations. The same ROI model and approach as before is applied. For each spectrum, we computed the likelihood ratio between a power-law and log parabolic power-law model. No significant ($>$3$\sigma$) preference is found for the log parabolic fit, and therefore a power-law is chosen. The number of spectral bins for the SEDs is chosen according to the time intervals and flux levels to assure at least four SED points for each of the three states described in Section~\ref{sec:spectral_analysis}. The obtained parameters are summarized in Table~\ref{tab:spec_par}.

\subsection{\textit{NuSTAR}}  \label{subsec:obs_analysis_nustar}

This work includes four multi-hour exposures from \textit{NuSTAR} \citep{2013ApJ...770..103H}. The \textit{NuSTAR} instrument consists of two co-aligned X-ray telescopes focusing on two independent focal plane modules, FPMA \& FPMB, and provides unprecedented sensitivity in the 3--79\,keV band. In this work, we analyze all \textit{NuSTAR} observations from 2022, which all took place in 2022-03: 09,  22, 24, 27 (MJD 59647, MJD 59660, MJD 59662, MJD 59665; observation ID 60701032002, 60502009002, 60502009004 \& 60702062004, respectively). 

The raw data are processed using the \textit{NuSTAR} Data Analysis Software (NuSTARDAS) package v.2.1.1 and CALDB version 20220912. The events are screened in the \texttt{nupipeline} process with the flags \texttt{tentacle=yes} \& \texttt{saamode=optimized} to remove potential background increase caused by the South Atlantic Anomaly passages. The source counts are obtained from a circular region centered around Mrk~501 with a radius of $\approx140''$. The background events are extracted from a source-free nearby circular region having the same radius. The spectra are then grouped with the \texttt{grppha} task to obtain at least 20 counts in each energy bin. \par 

For all exposures, the source spectra dominate over the background up to roughly $\approx 30$\,keV. Hence, we decide to quote fluxes only up to 30\,keV, and in two separate energy bands: 3--7\,keV and 7--30\,keV. The best-fit spectral parameters are obtained in the full \textit{NuSTAR} band-pass, 3--79\,keV, and averaged over the respective observations. We fit the spectra using \texttt{XSPEC} \citep{1996ASPC..101...17A} assuming a log-parabolic function with a normalization energy fixed to 1\,keV. A log-parabola model is significantly preferred over a simple power-law. Here, and for the rest of the X-ray analysis performed in this work, a photoelectric absorption component is added to the model assuming an equivalent hydrogen column density fixed to $N_{\rm H}=1.7\times 10^{20}$\,cm$^{-2}$ \citep{2016A&A...594A.116H}. The fluxes and spectral parameters are computed by fitting simultaneously FPMA \& FPMB. The cross-calibration factor between the two focal module planes is for all bins below 5\%, thus well within the expected systematics \citep{2015ApJS..220....8M}.\par 

For all \textit{NuSTAR} pointings, the variability is small and the fluxes vary at most by $\approx10\%$ within each observation. Thus, the fluxes and spectral parameters are averaged over the entire exposure time from each observation.

\subsection{\textit{IXPE}} 
The \textit{IXPE} telescope \citep{2022HEAD...1930101W} was launched in December 2021 and measures polarization in the 2--8\,keV regime. In this work, we use the first three pointing conducted on Mrk\,501. The first two took place from 2022-03-08 to 2022-03-10 (MJD 59646 to MJD 59648) and from 2022-03-27 to 2022-03-29 (MJD 59665 to MJD 59667) and the results are published in \citet{2022Natur.611..677L}, from which we extracted the flux and polarization values from Table 1 and 2.
A third pointing took place from 2022-07-09 to 2022-07-12 (MJD 59769 to MJD 59772) with the results published in \citet{2023_IXPE_Mrk501}. We extracted the polarization values from Table 2 of this paper and integrated the spectral log parabola model stated to obtain the corresponding flux values. 
During all pointings, no variability is observed in the polarization degree or angle, and the polarization and flux values are obtained averaged over the full exposures. For more details on the observations, data reduction, etc, see \citet{2022Natur.611..677L} and \citet{2023_IXPE_Mrk501}.

\subsection{\textit{Swift}-XRT}  \label{subsec:obs_analysis_swift_xrt}
Simultaneous to the MAGIC observations, several pointing of the XRT \citep[X-ray Telescope;][]{2005SSRv..120..165B} on board of \textit{Swift}
\citep[\textit{Neil Gehrels Swift Gamma-ray Burst Observatory};][]{2004ApJ...611.1005G} were organized. 
Data were taken both in the Windowed Timing (WT) and Photon Counting (PC) readout mode and processed with the XRTDAS software package (v3.7.0), that was developed by the ASI Space Science Data Center\footnote{\url{https://www.ssdc.asi.it/}} (SSDC), and released in the HEASoft package (v.6.30.1) by NASA High Energy Astrophysics Archive Research Center (HEASARC). For calibrating and cleaning the events, calibration files from \textit{Swift}-XRT CALDB (version 20210915) are processed within the \texttt{xrtpipeline}. 

For each observation, we extract the X-ray spectrum from the summed cleaned event file. In WT readout mode, the spectral analysis is performed after selecting events from a circle centered at the source position with a radius of 20 pixels ($\sim46$\arcsec; $\sim$90\% of the PSF). In the case of data taken in the PC mode, they are affected by pile-up in the inner part of the PSF. Therefore events from the most inner region around the source position within a radius of 4-6 pixels are excluded and an outer radius of 30 pixels is adopted to select signal events. We then use the shape of the \textit{Swift}-XRT PSF to calculate the incident X-ray flux\footnote{This is the standard pile-up correction procedure developed by the \textit{Swift}-XRT team and more details can be found here: \url{https://www.swift.ac.uk/analysis/xrt/pileup.php}}. To estimate the background, we use nearby circular regions with radii of 20 and 40 pixels for WT and PC data. The task \texttt{xrtmkarf} is used to produce ancillary response files (ARFs) including corrections for PSF losses and CCD defects using the cumulative exposure map. The X-ray spectra in the 0.3--10\,keV range are binned with \texttt{grppha} ensuring a minimum of 20 counts per bin. Afterwards, we model them in \texttt{XSPEC} using power-law and log-parabola models. Energy fluxes are computed in the 0.3--2\,keV and 2--10\,keV bands.

\subsection{\textit{Swift}-UVOT}  \label{subsec:obs_analysis_swift_uvot}
Simultaneous to the XRT pointings, the Ultraviolet/Optical Telescope \citep[UVOT; ][]{Roming05} also observed in the optical/UV wavebands. In this work, we considered images acquired with the three UV filters, W1 (2600\,\AA), M2 (2246\,\AA) and W2 (1928\,\AA), which are less affected by the host galaxy emission than the optical ones. The standard UVOT software within the HEAsoft package (v.6.23) is used to perform aperture photometry for all filters and the calibration is performed using CALDB (20201026). We review images to avoid unstable attitudes, then following \citet{Poole08}, we use a circular aperture of 5\arcsec~to extract source counts and an annular aperture with 26\arcsec~(34\arcsec) for the inner (outer) radii to estimate background counts. Using the standard zero points reported by \citet{brev11}, the count rates are converted to fluxes and then dereddened considering the $E(B-V)$ value 0.017 \citep{schlegel1998,Schlafly_2011} for the UVOT filters effective wavelengths and the mean galactic interstellar extinction curve from \citet{Fitzpatrick1999}.

\subsection{Optical}  \label{subsec:obs_analysis_optical}

Optical data including flux and polarization measurements were obtained from the 2.5m Nordic Optical Telescope (NOT), the 2.2m telescope of the Calar Alto Observatory as part of the Monitoring AGN with Polarimetry at the Calar Alto Telescopes (MAPCAT)\footnote{\href{https://home.iaa.csic.es/~iagudo/_iagudo/MAPCAT.html}{https://home.iaa.csic.es/~iagudo/\_iagudo/MAPCAT.html}}, the 1.5m (T150) and the 0.9 m (T090) telescopes at the Sierra Nevada Observatory, the 1.3m RoboPol telescope \citep[Skinakas observatory, Greece;][]{2014MNRAS.442.1706K, 2019MNRAS.485.2355R}, the 1.5m KANATA (Higashi-Hiroshima observatory, Japan) telescopes and network of robotic telescopes BOOTES \citep[Burst Observer and Optical Transient Exploring System;][]{1999A&AS..138..583C}. Except for BOOTES, which uses SDSS filters, Johnson-Cousins filters are used. To convert magnitudes between the different filters \citet{2005AAS...20713308L} is used.

Additionally, Tuorla blazar monitoring data of Mrk\,501 for the period between 2022-03-01 and 2022-07-19 (MJD 59639 to MJD 59779) were obtained using the 80\,cm Joan Oró Telescope (TJO) at Montsec Observatory, Spain. The observations were made in the Cousins R-filter. The flux density values are obtained following standard photometry analysis \citep{Nilsson_2018} using a signal extraction aperture of 7.5". Additionally, to correct for the flux density contamination coming from the host galaxy and nearby sources a 12.3 mJy subtraction was implemented following \citet{Nilsson_2007}.

Flux and polarization degree values were corrected for the host galaxy contribution following \citet{Nilsson_2007}, \citet{Weaver2020} or \citet{2016A&A...596A..78H} taking into account the apertures of the telescopes and the seeing conditions. Additionally, the galaxy extinction correction of 0.041 mag was implemented following \citet{Schlafly_2011} for the flux values. 

To also cover the second \textit{IXPE} pointing, we added the data taken by the  70\,cm AZT-8 telescope of the Crimean Astrophysical Observatory\footnote{In 1991, Ukraine with the Crimean peninsula became an independent state. While the Crimean Astrophysical Observatory became Ukrainian, the AZT-8 telescope located there continued to be operated jointly by the Crimean Observatory and by the St. Petersburg group.}, which took both photometric and polarimetric data with the Cousins R-filter from 2022-03-25 to 2022-03-28 (MJD 59663 to MJD 59666) published in \citet{2022Natur.611..677L} and applied the same host galaxy and the Galactic extinction correction to the optical flux. For the polarization degree, values that had already been corrected are used.

\subsection{Radio}  \label{subsec:obs_analysis_radio}

In the radio band, observations were taken on Mrk\,501 by the single-dish telescopes at the Owens Valley Radio Observatory (OVRO) operating at 15 GHz, the Mets\"{a}hovi Radio Observatory at 37\,GHz, the Institut de Radioastronomie Millimetrique (IRAM) at 86 GHz and 230\,GHz and the Submillimeter Array (SMA) at 226\,GHz.

The data from OVRO were taken with the 40\,m telescope blazar monitoring program using a central frequency of 15\,GHz and 3\,GHz bandwidth as detailed in \citet{Richards_2011}. 
37\,GHz data were taken by the Mets\"ahovi Observatory and analyzed according to \citet{1998A&AS..132..305T} using NGC\,7027, 3C\,274 and 3C\,84 as secondary calibrators. Estimates on the error of the flux density contain the contribution from the measurement root mean square and the uncertainty of the absolute calibration.  

Additional flux, but also polarization data, in the radio band were collected by the IRAM 30\,m telescope under the Polarimetric Monitoring of AGN at Millimeter Wavelengths (POLAMI) program\footnote{\url{https://polami.iaa.es/}} \citep{2018MNRAS.473.1850A, 2018MNRAS.474.1427A} in the 3.5\,mm (86.24\,GHz) and 1.3\,mm (230\,GHz) wavebands.
The XPOL procedure \citep{2008PASP..120..777T} was used to simultaneously measure the four Stokes parameters (I, Q, U and V) at the two observing bands. The POLAMI pipeline described in \citet{2018MNRAS.474.1427A} was adopted to reduce and calibrate the data.

The Submillimeter Array \citep[SMA;][]{Ho2004} was used to obtain polarimetric millimeter radio measurements at $\sim$1.3~mm ($\sim226$~GHz) within the framework of the SMAPOL (SMA Monitoring of AGNs with POLarization) program. SMAPOL follows the polarization evolution of forty $\gamma$-ray loud blazars, including Mrk\,501, on a bi-weekly cadence as well as other sources in a target-of-opportunity (ToO) mode. The Mrk\,501 observations reported here were conducted in June and July, 2022.
The SMA observations use two orthogonally polarized receivers, tuned to the same frequency range in the full polarization mode, and use the SWARM correlator \citep{Primiani2016}. These receivers are inherently linearly polarized but are converted to circular using the quarter-wave plates of the SMA polarimeter \citep{Marrone2008}. The lower sideband (LSB) and upper sideband (USB) covered 209-221 and 229–241 GHz, respectively. Each sideband was divided into six chunks, with a bandwidth of 2 GHz, and a fixed channel width of 140 kHz. The SMA data were calibrated with the MIR software package \footnote{\href{https://lweb.cfa.harvard.edu/~cqi/mircook.html}{https://lweb.cfa.harvard.edu/~cqi/mircook.html}}. Instrumental polarization leakage was calibrated independently for USB and LSB using the MIRIAD task gpcal \citep{Sault1995} and removed from the data.  The polarized intensity, position angle, and polarization percentage were derived from the Stokes I, Q, and U visibilities.
MWC\,349\,A, Callisto and Titan were used for the total flux calibration and the calibrator 3C\,286, which has high linear polarization degree and stable polarization angle, was observed as a cross-check of the polarization calibration.

\section{Detailed VHE \& X-ray analysis during IXPE observations} 
\label{sec:ixpe_mwl}

\begin{figure*}
   \centering
   \includegraphics[width=1.85\columnwidth]{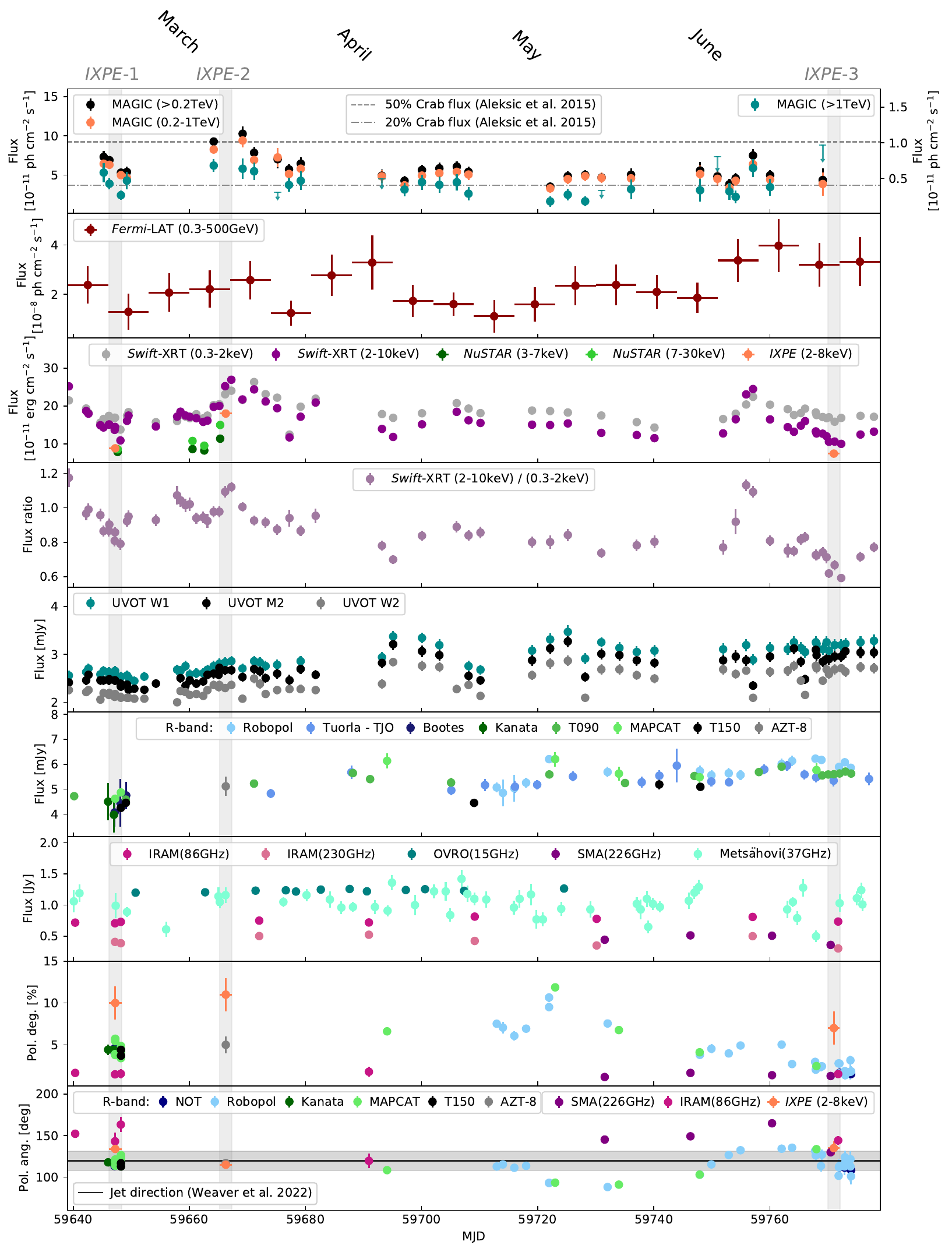}
   \caption{\small MWL light curve for Mrk\,501 between 2022-03-01 and 2022-07-19 (MJD 59639 to MJD 59779). The grey areas mark the IXPE observations taking place between 2022-03-08 to 2022-03-10 (MJD~59646 to MJD~59648), 2022-03-27 to 2022-03-29 (MJD~59665 to MJD~59667) and 2022-07-09 to 2022-07-12 (MJD~59769 to MJD~59772). Top to bottom: MAGIC fluxes in daily bins with the arrows additionally displaying the ULs (95\% confidence level) for the non-significant bins ($<$2\,$\sigma$); \textit{Fermi}-LAT fluxes in 7\,day bins; X-ray fluxes in daily bins including \textit{Swift}-XRT, \textit{NuSTAR} and IXPE; hardness ratio between the high- and low-energy fluxes of \textit{Swift}-XRT; \textit{Swift}-UVOT; Optical R-band data from Robopol, Calar Alto, T090, MAPCAT, T150, AZT-8 in daily bins; Radio data including OVRO, Mets\"ahovi, IRAM, and SMA; polarization degree \& polarization angle observations in the X-rays from \textit{IXPE}, the optical R-band from NOT, Robopol, Calar Alto, MAPCAT, T150 and AZT-8 and the radio band from IRAM and SMA. The \textit{IXPE} and AZT-8 data are taken from \citet{2022Natur.611..677L} and \citet{2023_IXPE_Mrk501}. In the polarization angle panel, we plot with a horizontal grey band the average direction angle determined at 43\,GHz by \citet{Weaver_2022}. See text in Section~\ref{sec:longterm_MWL} for further details.}
    \label{fig:MWL_LC}%
\end{figure*}

The MWL light curves from 2022-03-01 to 2022-07-19 (MJD 59639 to MJD 59779) are shown in Figure~\ref{fig:MWL_LC}. The three \textit{IXPE} observations conducted during this time range are highlighted with a grey band. Two observations took place from 2022-03-08 to 2022-03-10 (MJD 59646 to MJD 59648) and from 2022-03-27 to 2022-03-29 (MJD 59665 to MJD 59667). The third one took place from 2022-07-09 to 2022-07-12 (MJD 59769 to MJD 59772). In the following, those three periods will be referred to as \textit{IXPE}-1, \textit{IXPE}-2 and \textit{IXPE}-3, respectively. This section discusses the MWL states during these observations with a focus on the VHE and X-ray regimes.

\subsection{IXPE observations in March 2022}
\label{sec:march_2022_lc_description}

The first two \textit{IXPE} observations from 2022-03-08 to 2022-03-10 and 2022-03-27 to 2022-03-29 (MJD 59646 to MJD 59648 and MJD 59665 to MJD 59667), \textit{IXPE}-1 and \textit{IXPE}-2, were reported in \citet{2022Natur.611..677L}. The measured polarization degree in the 2--8\,keV band for those observations are $10\% \pm 2\%$ and $11\% \pm 2\%$, respectively, which is around 2 times higher than the one measured at optical wavelengths. The corresponding polarization angles, $134\pm 5^\circ$ and $115\pm 4^\circ$, match the optical and radio measurements and are in line with the radio jet direction of $119.7 \pm 11.8^\circ$ \citep[][depicted as a horizontal grey band in Figure~\ref{fig:MWL_LC}]{Weaver_2022}. The 2--8\,keV flux of the second pointing ($18.0\pm0.4\times 10^{-11}$ erg cm$^{-2}$ s$^{-1}$) is a factor of two higher than for the first one ($8.8\pm0.1\times 10^{-11}$ erg cm$^{-2}$ s$^{-1}$).

Close to the \textit{IXPE}-1 epoch, four MAGIC observations are available (see Figure~\ref{fig:MWL_LC}). Two observations took place strictly simultaneous to the \text{IXPE} pointing, while the two others happened one day before and after. The nightly light curve is consistent with a constant flux hypothesis, with  a corresponding average flux of $6.07 \pm 0.29\times 10^{-11}$\,ph\,cm$^{-2}$\,s$^{-1}$ ($\sim$30\% C.U.) above 200\,GeV.

Due to bad weather, no MAGIC simultaneous data are available during \textit{IXPE}-2. However, one observation took place one day before the \textit{IXPE} window and another one two days after (Figure~\ref{fig:MWL_LC}). Between those two MAGIC observations encompassing the \textit{IXPE}-2 epoch, the MWL fluxes show little variability, in particular in the X-ray band (see Figure~\ref{fig:MWL_LC}), which is typically well correlated with the VHE emission \citep{Mrk501_MAGIC_2014,Mrk501_MAGIC_2013,Mrk501_MAGIC_2012, 2023ApJS..266...37A}. We thus average the two MAGIC exposures to obtain a reasonable approximation of the VHE state simultaneous to the \textit{IXPE}-2 exposure. The resulting average flux is $9.53 \pm 0.44\times 10^{-11} $\,ph\,cm$^{-2}$\,s$^{-1}$ above 200\,GeV ($\sim$50\% C.U.), which is $\sim40\%$ higher than during \textit{IXPE}-1. Overall, the \textit{IXPE}-1 and \textit{IXPE}-2 VHE levels are within the typical dynamical flux range for Mrk~501 as they remain close to the average state \citep{Abdo_2011} and are persistently above its low state flux observed from 2017 to 2019 \citep{2023ApJS..266...37A}.  

Similar to the VHE band, the \textit{Swift}-XRT and \textit{IXPE} measurements also reveal X-ray flux levels that are around a factor of 2 higher during the \textit{IXPE}-2 epoch compared to \textit{IXPE}-1. On the other hand, while the VHE emission is around the average source activity, the fluxes in the \textit{Swift}-XRT bands are systematically above average. The 2--10\,keV fluxes lie between $\approx 10^{-10}$\,erg\,cm$^{-2}$\,s$^{-1}$ and $\approx 3 \times 10^{-10}$\,erg\,cm$^{-2}$\,s$^{-1}$, which is significantly higher than the levels reported in previous works that studied the source during an average VHE activity such as a flux of $7.8 \pm 0.2 \times 10^{-11}$\,erg\,cm$^{-2}$\,s$^{-1}$ in 2009 \citep[][]{Abdo_2011} or a weighted mean of $4.24 \pm 0.01 \times 10^{-11}$\,erg\,cm$^{-2}$\,s$^{-1}$ for 12-year flux data set from 2008 to 2020 \citep{2023ApJS..266...37A}. Regarding the variability, we find that the \textit{Swift}-XRT data display only slight flux changes (at the level of $\sim$10-30\%) within each \textit{IXPE} window. No intra-night variability is detected, in either the VHE or the X-ray regimes.

Additionally, four \textit{NuSTAR} pointings were conducted during 2022-03-09, 2022-03-22, 2022-03-24 and 2022-03-27 (MJD 59647, MJD 59660, MJD 59662, MJD 59665). One simultaneous to each \textit{IXPE}-1 and \textit{IXPE}-2 window, while two  other observations happened shortly before the \textit{IXPE}-2 pointing. The flux ranges from $7.8$ to $11.4 \times 10^{-11}$ erg cm$^{-2}$ s$^{-1}$ in the 3--7\,keV band, which is at least a factor of 5 higher than during the previous low-activity measurements in 2017-2018 \citep{2023ApJS..266...37A}, but similar to the observations reported in 2013 \citep{Mrk501_MAGIC_2013} for Mrk\,501. The \textit{NuSTAR} exposures, which are multi-hour long, offer the possibility to search for short timescale variability. No strong X-ray variability is found on hourly timescale, and throughout each exposure the flux varies by at most 10\%.

In the radio, optical, \textit{Swift}-UVOT and \textit{Fermi}-LAT wavebands, the \textit{IXPE}-1 and \textit{IXPE}-2 epochs are in a rather averaged and stable state \citep{Abdo_2011}, without significant flux variations on daily timescales.

\begin{table*}
\centering
\caption{Parameters for the VHE and X-ray observations around the three IXPE pointings.}
\begin{tabular}{ c c || c | c | c}     
& Observations & \textit{IXPE}-1 & \textit{IXPE}-2 & \textit{IXPE}-3 \\
\hline
& MJD & 59646.11 to 59648.35 & 59665.24 to 59667.30 &  59769.97 to 59772.04 \\
 \hline  \hline
 \multicolumn{1}{c|}{\multirow{4}{*}{\rotatebox{90}{MAGIC}}}  & MJD & 59645.24 to 59649.25 & 59664.19 to 59669.19  &  59769.10 to 59769.11  \\
 \multicolumn{1}{c|}{} & Flux & $0.45 \pm 0.03$   &  $0.73  \pm 0.04$ &  $0.32  \pm 0.01$ \\
 \multicolumn{1}{c|}{} & $\alpha$ & 2.47 $\pm$ 0.04 &  2.44 $\pm$ 0.04 & 2.50 $\pm$ 0.24 \\
 \multicolumn{1}{c|}{} & $\chi^{2}$/d.o.f. & 22.15/14 &  22.15/14 & 5.10/4 \\
\hline
 \multicolumn{1}{c|}{\multirow{3}{*}{\rotatebox{90}{LAT}}}  & MJD & 59640.23 to 59654.23 & 59659.27 to 59673.27  &  59764.01 to 59778.01 \\
 \multicolumn{1}{c|}{} & Flux & $2.55 \pm 1.24$ &  $1.74 \pm 0.74$ & $1.00 \pm 0.31$  \\
 \multicolumn{1}{c|}{} & $\alpha$ & 1.60 $\pm$ 0.13 &  1.80 $\pm$ 0.14 & 2.10 $\pm$ 0.17 \\
\hline
 \multicolumn{1}{c|}{\multirow{4}{*}{\rotatebox{90}{\textit{NuSTAR}}}}  & MJD & 59647.42 to 59647.88 & 59665.06 to 59665.55 &  (...) \\
 \multicolumn{1}{c|}{} & Flux & $0.784\pm0.003$ & $1.135\pm0.004$ & (...) \\
 \multicolumn{1}{c|}{} & $\alpha$ & $1.851\pm0.007$ & $1.684\pm0.005$  & (...) \\
 \multicolumn{1}{c|}{} & $\chi^{2}$/d.o.f. & 951.6/925 &  1150.2/1123& (...) \\
\hline
 \multicolumn{1}{c|}{\multirow{8}{*}{\rotatebox{90}{\textit{Swift}-XRT}}}  & MJD & 
 \multirow{4}{*} {\begin{tabular}{@{}ccc@{}} 59646.13 & 59646.21 & 59647.13 \\
  $1.51 {}^{+0.06}_{-0.05}$&$1.51 {}^{+0.06}_{-0.05}$&$1.37 {}^{+0.05}_{-0.05}$ \\
   $1.95{}^{+0.02}_{-0.02}$& $1.95{}^{+0.02}_{-0.02}$& $2.01{}^{+0.02}_{-0.02}$ \\
  258.5/212&  268.28/219& 221.22/217  
  \end{tabular}}
 & \multirow{4}{*} {\begin{tabular}{@{}ccc@{}} 59664.18 & 59665.16 & 59666.16 \\
  $1.98 {}^{+0.06}_{-0.06}$&$2.00 {}^{+0.06}_{-0.06}$&$2.53 {}^{+0.07}_{-0.07}$ \\
 $1.90 {}^{+0.02}_{-0.02}$&$1.90 {}^{+0.02}_{-0.02}$&$1.84 {}^{+0.02}_{-0.02}$  \\
 331.22/253  & 271.97/241 & 266.25/274 
   \end{tabular}}
 &  \multirow{4}{*} {\begin{tabular}{@{}cc@{}} 59770.16 & 59771.15  \\
$1.05{}^{+0.04}_{-0.04}$&$1.06 {}^{+0.05}_{-0.04}$ \\
$2.16 {}^{+0.02}_{-0.02}$&$2.12 {}^{+0.02}_{-0.02}$  \\
190.39/195 &  192.28/181 
   \end{tabular}} \\
\multicolumn{1}{c|}{}& Flux &  &  &    \\
   \multicolumn{1}{c|}{}& $\alpha$ &  &  &    \\
\multicolumn{1}{c|}{}& $\chi^{2}$/d.o.f. &  &  &    \\
 \multicolumn{1}{c|}{} & MJD & 
 \multirow{4}{*} { \begin{tabular}{@{}cc@{}} 59647.20 & 59648.12 \\
  $1.45{}^{+0.05}_{-0.04}$& $1.09 {}^{+0.05}_{-0.04}$ \\
   $1.98{}^{+0.02}_{-0.02}$& $2.02 {}^{+0.02}_{-0.02}$ \\
  232.89/220 & 216.78/197 
  \end{tabular}}
 & \multirow{4}{*} { \begin{tabular}{@{}cc@{}} 59667.23 & 59669.14\\
  $2.70 {}^{+0.07}_{-0.07}$& $2.17 {}^{+0.05}_{-0.06}$\\
 $1.82 {}^{+0.01}_{-0.01}$& $1.88 {}^{+0.01}_{-0.01}$   \\
292.01/289 & 307.98/278 
   \end{tabular}}
 &  \\
  \multicolumn{1}{c|}{} & Flux &  &  & (...)   \\
  \multicolumn{1}{c|}{} & $\alpha$ &  & &  (...)  \\
 \multicolumn{1}{c|}{}  & $\chi^{2}$/d.o.f. &  & & (...) \\

\end{tabular}
\tablefoot{Stated are the MJD of the observations, the flux values in $10^{-10}$\,erg\,cm$^{-2}$\,s$^{-1}$ (for MAGIC with $E>200$\,GeV; for \textit{Fermi}-LAT from $0.3-300$\,GeV; for XRT from $2-10$\,keV; from $3-7$\,keV for \textit{NuSTAR}), the spectral index $\alpha$ and the $\chi^{2}$/d.o.f. for the spectral fits. For all wavebands, a simple power law is used as a spectral model (with a normalization energy of 300\,GeV for MAGIC, 1\,GeV for \textit{Fermi}-LAT and 1\,keV for \textit{Swift}-XRT) because there is no significant preference for a log-parabola function, except for the \textit{NuSTAR} data. The \textit{NuSTAR} fits are performed using a log-parabola model with a curvature parameter fixed to 0.3 (and normalization energy at 1\,keV).}
\label{tab:spec_par}
\end{table*}

\begin{table}
\centering
\caption{Peak frequencies, $\nu_\text{s}$ and $\nu_\text{IC}$, and Compton dominance (CD) for the different SEDs shown in Fig.~\ref{fig:SED} extracted from the maxima of the phenomenological description of \citet{blazar_seq}.}
\setlength{\tabcolsep}{0.4em} 
\begin{tabular}{ l | c c c}  
\multirow{2}{*}{States} & $\nu_\text{s}$ & $\nu_\text{IC}$ & CD\\
 &  [Hz] & [Hz] & \\
\hline \hline
\textit{IXPE}-1$_\text{pheno}$ &  $5.4 \pm 0.2 \times10^{17} $ & $2.0 \pm 0.4 \times10^{25}$ & 0.30 $\pm$ 0.07\\
\textit{IXPE}-2$_\text{pheno}$ &  $7.9 \pm 0.6 \times10^{17}$ & $3.0 \pm 0.5\times10^{25}$ & 0.28 $\pm$ 0.06 \\
\textit{IXPE}-3$_\text{pheno}$ &  $1.0 \pm 0.1 \times10^{17}$ & $4.5\pm 5.9\times10^{24}$ & 0.20 $\pm$ 0.27 \\
\hline
Typical$_\text{pheno}$ &  $2.9 \pm 0.1 \times10^{17} $ & $4.6 \pm 0.8 \times10^{25}$ & 0.49 $\pm$ 0.10\\
Low$_\text{pheno}$ &  $1.3 \pm 0.1 \times10^{16} $ & $1.8 \pm 0.4 \times10^{24}$ & 0.26 $\pm$ 0.10\\
\end{tabular}
\label{tab:peak_freq}
\end{table}

\subsection{IXPE observation in July 2022}
\label{sec:july_2022_lc_description}
The third \textit{IXPE} pointing, \textit{IXPE}-3, took place from 2022-07-09 to 2022-07-12 (MJD 59769 to MJD 59772). It is characterized by a slightly lower flux level ($7.4\pm0.4\times 10^{-11}$ erg cm$^{-2}$ s$^{-1}$ in the 2--8\,keV band) and polarization degree ($7\pm2$\%) in the X-rays than for the \textit{IXPE}-1 \& \textit{IXPE}-2 pointings \citep{2023_IXPE_Mrk501}. For MAGIC, no strictly simultaneous data are available, but a single observation took place less than one day before \textit{IXPE}-3. It shows a lower flux state compared to the ones in March with a flux of $4.36  \pm 1.45\times 10^{-11}$\,ph\,cm$^{-2}$\,s$^{-1}$ above 200\,GeV ($\sim$20\% C.U.). The X-ray fluxes of \textit{Swift}-XRT are lower than during \textit{IXPE}-2, although comparable to the ones measured during \textit{IXPE}-1.

\subsection{Spectral evolution during IXPE campaigns}
\label{sec:spectral_analysis}

\begin{figure*}
\centering
  \resizebox{0.8\hsize}{!}{\includegraphics{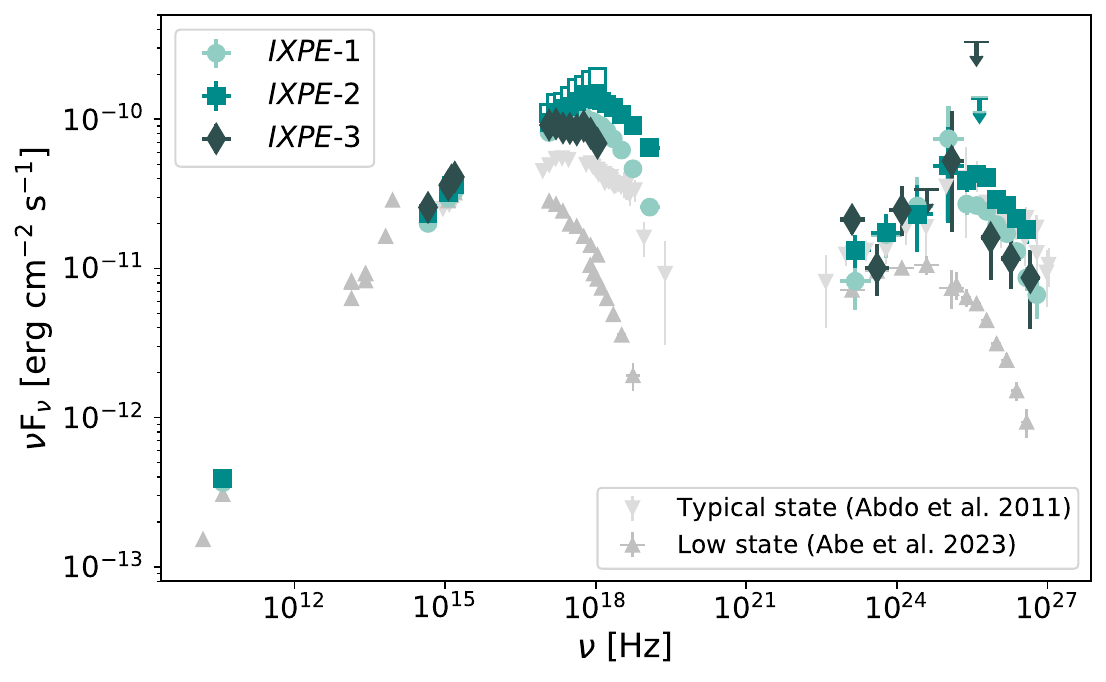}}
  \caption{Broad-band SED for the three IXPE windows as described in Section~\ref{sec:spectral_analysis}. For the \textit{IXPE}-2, the X-ray spectrum corresponding to the highest flux state in the time window is shown with open markers. The typical state \citep{Abdo_2011} and low state \citep{2023ApJS..266...37A} of Mrk\,501 are shown in grey for comparison.}
  \label{fig:SED}
\end{figure*}

Table~\ref{tab:spec_par} reports the fluxes and spectral parameters during each \textit{IXPE} epoch for the MAGIC, \textit{Fermi}-LAT, \textit{NuSTAR} and \textit{Swift}-XRT observations. For all instruments except \textit{NuSTAR}, we fit a simple power-law model given the absence of significant evidence for spectral curvature. The \textit{NuSTAR} spectra show a significant curvature and are thus fitted with a log-parabola model, $\frac{dN}{dE} \propto \left(E/E_{norm}\right)^{-\alpha-\beta \log(E/E_{norm})}$ (with a normalization energy of $E_{norm}=1$\,keV). Furthermore, we fix the curvature parameter to $\beta=0.3$ (average value of the pointings) to remove any correlation between $\alpha$ and $\beta$, and obtain a better assessment of the spectral hardness evolution. For each instrument, we state in the corresponding panels the time ranges over which the spectra are computed. For \textit{Swift}-XRT, the best-fit parameters are shown for each observation separately given the indication of daily spectral evolution.

Comparing the three \textit{IXPE} epochs, one can see the higher and harder emission for \textit{IXPE}-2, in particular in the X-ray band. In fact, \textit{IXPE}-2 is characterized by one of the hardest X-ray states from the campaign discussed in this work ($\alpha<1.9$). Such a hard state is not evident in the VHE band, which in fact shows a constant power-law index between the three epochs within uncertainties ($\alpha\sim2.4-2.5$, see Table~\ref{tab:spec_par}). The \textit{Swift}-XRT spectra show a ''harder when brighter'' trend that is also observed when comparing the \textit{NuSTAR} pointings. In the \textit{Fermi}-LAT band, the spectral index softens along the three \textit{IXPE} time windows leading to a decrease in energy flux despite increasing photon flux levels.

We construct MWL SEDs for all pointings. Except for the $\gamma$-ray regime, we used the weighted average of all data points in the corresponding \textit{IXPE} time windows as stated in the first row of Table~\ref{tab:spec_par}. The only exception is made for the \textit{IXPE}-2 spectrum, for which the dedicated \textit{NuSTAR} observation started shortly before the \textit{IXPE} time window. Therefore, the \textit{NuSTAR} observational window on 2022-03-27 (MJD~59665) is used to extract the weighted averaged SED points for the radio, UV and X-ray bands. For the optical regime, the data simultaneous to \textit{IXPE}-2 from \citet{2022Natur.611..677L} are used because they are the only  data available around this time window. This ensures a smooth characterization between the two X-ray instruments and therefore of the synchrotron peak. For comparison the \textit{Swift}-XRT spectrum corresponding to the highest flux level during the \textit{IXPE}-2 window is shown with open markers in Figure~\ref{fig:SED}. For \textit{Fermi}-LAT a time window of two weeks centered on the \textit{IXPE} epochs is used to construct the SEDs owing to the low variability in the band and the lower instrument sensitivity, which requires longer observation times. For MAGIC, simultaneous data to the \textit{IXPE} observations are only available for \textit{IXPE}-1. We used all simultaneous nights and also the two surrounding ones to construct the \textit{IXPE}-1 MAGIC SED because no flux or spectral variations are seen between the different days. Due to the availability of VHE observations as discussed in Section~\ref{sec:march_2022_lc_description} and Section~\ref{sec:july_2022_lc_description} and the absence of strong spectral and flux variability at other wavebands, the observations surrounding the \textit{IXPE} windows are used to construct the \textit{IXPE}-2 and \textit{IXPE}-3 VHE SEDs.

The resulting SEDs are shown in Figure~\ref{fig:SED}. 
For comparison purposes, we plot in light gray the average source state from \citet{Abdo_2011}, and in dark gray the low activity discussed in \citet{2023ApJS..266...37A}. As anticipated in Section~\ref{sec:march_2022_lc_description}, Figure~\ref{fig:SED} highlights a drop in the Compton dominance (CD; the ratio between the $\nu F_\nu$ values at the high-energy and low-energy peaks) with respect to the average activity (light gray points). The entire $\gamma$-ray SED component remains close to the average activity, while the X-ray emission is significantly higher.

Using the phenomenological model of \citet{blazar_seq}, we quantify the peak frequencies and CD for each epoch. The same phenomenological model is fitted to the average and low states from \citet{Abdo_2011} and \citet{2023ApJS..266...37A} for comparison purposes. The results are shown in Table~\ref{tab:peak_freq}. 
The obtained best-fit value place both \textit{IXPE}-1 and \textit{IXPE}-2 in the EHSP \citep{2001A&A...371..512C, 2020NatAs...4..124B} family with $\nu_\text{s}>2.4\times 10^{17}$\,Hz ($\approx1$\,keV). 

For the third pointing, one can clearly see that the synchrotron peak moved to lower frequency, into the HSP state as verified by the phenomenological fits (see Table~\ref{tab:peak_freq}) yielding a $\nu_\text{s}$ of $\sim1\times10^{17}$\,Hz. While the MAGIC spectrum shows only a change in amplitude (without spectral evolution), the \textit{Fermi}-LAT spectrum hints towards softer emission, leading to a shift of the high-energy peak towards lower frequencies by an order of magnitude. 

For all three spectra, the fit of the phenomenological model confirms a general lower CD (for \textit{IXPE}-1 and \textit{IXPE}-2 $\sim$0.3, for \textit{IXPE}-3 0.2 using the phenomenological model or 0.3 using the SSC model described in Section~\ref{sec:modelling}) than for the average state of Mrk\,501, which is determined to be around $\sim$0.5.

\subsection{Evolution of the polarization degree between the IXPE epochs}
\label{sec:pol_vs_freq}

\begin{figure}
\centering
  \resizebox{0.9\hsize}{!}{\includegraphics{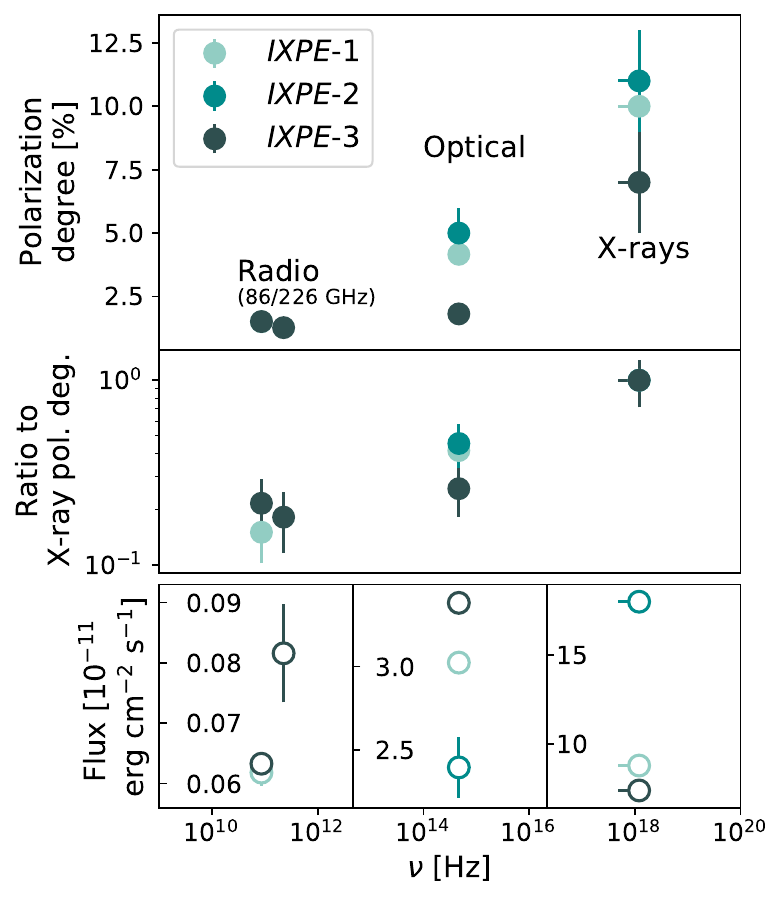}}
  \caption{Top: Multi-wavelength polarization degree as a function of frequency for the three \textit{IXPE} observations. For the radio and optical data, weighted averages over the \textit{IXPE} time windows are used. Middle: Ratio of the frequency dependent polarization degree to the corresponding X-ray polarization degree. Bottom: Flux values associated with the polarization values presented above.}
  \label{fig:pol_deg_evol}
\end{figure}

Figure~\ref{fig:pol_deg_evol} shows the polarization degree of all three \textit{IXPE} pointings in different wavebands, from the radio to the X-ray band. Additionally, we show in the middle panel the ratio of the polarization degree to the one observed in the X-rays. For all three epochs, the polarization degree increases with frequency showing a ratio that is slightly more than two between the X-ray and optical regimes. This behavior was first reported for the \textit{IXPE}-1 and \textit{IXPE}-2 epochs by \citet{2022Natur.611..677L}. \textit{IXPE}-3 follows the same trend, but it is interesting to see that while the absolute polarization degree drops with respect to \textit{IXPE}-1 \& \textit{IXPE}-2 in optical and X-rays, the ratio is roughly conserved and stable over time (the derived ratios remain consistent within $\sim1\sigma$). \par

As described in \citet{2022Natur.611..677L} and \citet{2022ApJ...938L...7D}, the increase of the polarization degree with energy, the absence of fast variability in the polarization behavior as well as the alignment of the radio jet with the polarization angle in optical and X-rays strongly suggest that electrons are accelerated in shocks. The most energetic electrons (those emitting X-ray photons) are confined nearby the shock front, where the magnetic field is more ordered, leading to a stronger polarization degree. The electrons subsequently cool, emit lower energy photons, and advect (or diffuse) towards larger regions in which the degree of magnetic field ordering drops significantly. Such energy-stratified scenario qualitatively explains the broad-band polarization degree evolution \citep{2020MNRAS.498..599T}, and will be considered for the MWL modeling of the SEDs presented in Section~\ref{sec:modelling}. 

The lower panel of Figure~\ref{fig:pol_deg_evol} shows the average flux states for the three epochs in each energy bands. No obvious coherent trend is apparent between the evolution of the polarization degree and the flux level. In fact, while in the optical the polarization degree tends to be anti-correlated with the flux, the X-ray hints towards an opposite trend.

\section{Long-term multi-wavelength evolution}
\label{sec:longterm_MWL}

\begin{figure}
\centering
  \resizebox{\hsize}{!}{\includegraphics{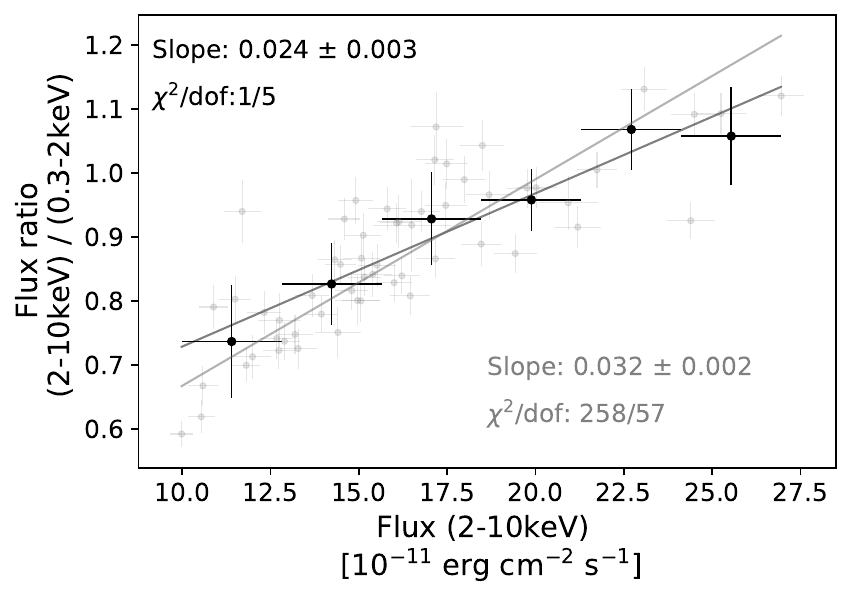}}
  \caption{\textit{Swift}-XRT hardness ratio (flux ratio of the 2--10\,keV range over the 0.3--2\,keV band) compared to flux at 2--10\,keV for the light curves shown in Figure~\ref{fig:MWL_LC}. The grey data points are from individual measurements, while the black points are the hardness ratio binned in flux.}
  \label{fig:xrt_hardness}
\end{figure}

\begin{figure}
\centering
  \resizebox{\hsize}{!}{\includegraphics{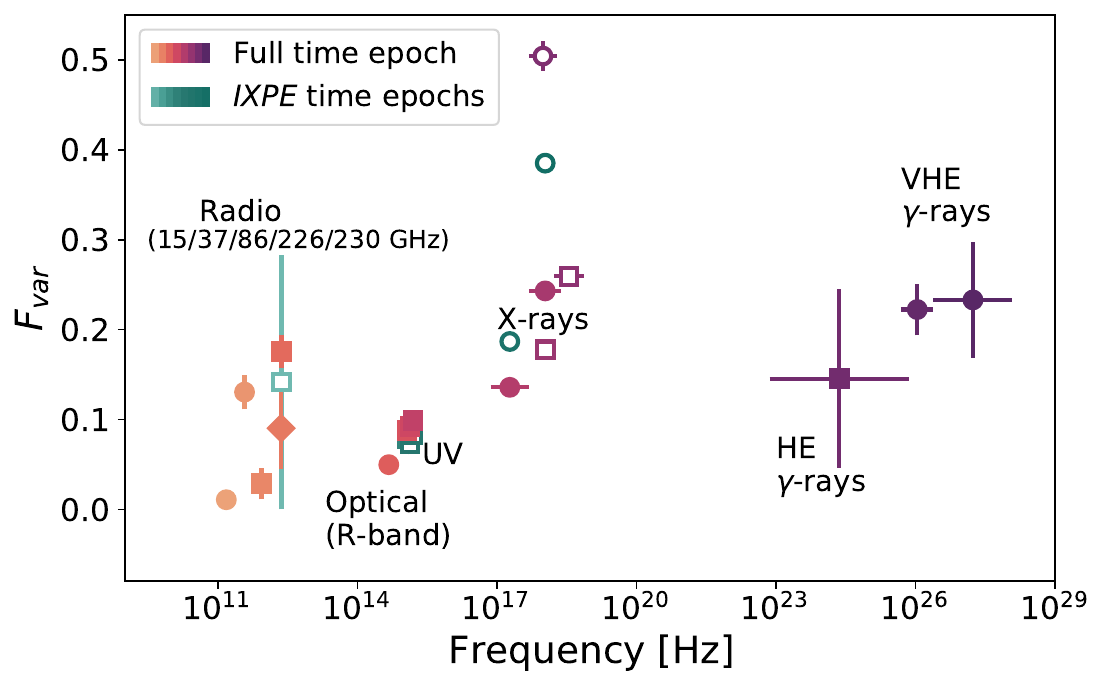}}
  \caption{Fractional variability for the light curves displayed in Figure~\ref{fig:MWL_LC}. For \textit{Fermi}-LAT 7 day bins and for MAGIC nightly bins are used, for all other wavebands the single observations without further binning are used. For the radio data, frequencies are stated to distinguish the different data sets (from left to right: the circle orange markers depict OVRO at 15\,GHz and Metsähovi at 37\,GHz, the square lighter orange marker the IRAM data at 86\,GHz, the diamond orange marker SMA at 226~GHz and the square darker orange and open turquoise marker IRAM at 230\,GHz). The violet open markers refer to the \textit{NuSTAR} (square) and \textit{IXPE} (round) observations, which only consist of 2-3 measurements and therefore far less data points than considered for the full time epoch. Due to the different sampling, not all instruments are directly comparable with each other.}
  \label{fig:fvar}
\end{figure}

This section discusses the MWL evolution throughout the entire period between 2022-03-01 to 2022-07-19 (MJD 59639 to MJD 59779) in order to provide a broader overview of the source evolution  between the several \textit{IXPE} epochs.\par 

From Figure~\ref{fig:MWL_LC}, the VHE flux shows no significant outbursts and varies around $\sim$20--50\% that of the Crab Nebula, which is close to the typical state of Mrk~501 \citep{Abdo_2011}. The low (0.2--1\,TeV) and high ($>1$\,TeV) energy ranges follow similar variability patterns. No ''harder when brighter'' trend is found, which however might be rooted in the lower instrumental sensitivity of MAGIC to unveil slight spectral changes compared to e.g. X-ray instruments.\par 

The \textit{Swift}-XRT bands reveal clear hardening with increasing flux. In Figure~\ref{fig:xrt_hardness}, we show the hardness ratio (defined as the ratio of the 2--10\,keV flux to the one in the 0.3--2\,keV band) as a function of the 2--10\,keV flux. The grey data points are from individual measurements, while the black points are the hardness ratio binned in flux. The error bars represents the standard deviation of the hardness ratio in each bin. A positive slope of $0.032\pm0.002$ is found when fitting the individual data points with a linear function. For the hardness ratio binned in flux, the best fit slope is comparable, $0.024\pm0.003$. These slopes are significantly lower than what was obtained in \citet{2023ApJS..266...37A} during a much lower activity of Mrk~501, possibly indicating a saturation of the hardness ratio above a certain flux level similarly to what was noted in Mrk~421 in \citet{2021MNRAS.504.1427A}.

For the UV band, we see a change in flux happening around April slightly increasing the average UV flux level, which can be explained by the shift of the synchrotron peak to lower frequencies. The latter explanation is also suggested by comparing the three \textit{IXPE} SEDs in Figure~\ref{fig:SED}, which unveil a mild increase in the optical/UV during the \textit{IXPE}-3 period that displays a shift of the synchrotron component towards lower energies.

We quantify the variability strength throughout the spectrum using the fractional variability ($F_{var}$) defined following Eq. 10 in \citet{2003MNRAS_frac_var}

\begin{equation}
    F_{var} = \sqrt{\frac{S^2 - <\sigma_{err}^2>}{<F_\gamma>^2}}
\end{equation}
with the variance of the signal $S^2$, the mean square error of the measurement uncertainties $<\sigma_{err}^2>$ and the arithmetic mean of the measured flux $<F_\gamma>^2$. Its uncertainty is computed following \citet{2008MNRAS.389.1427P}:
\begin{equation}
    \Delta F_{var} = \sqrt{F_{var}^2 + err(\sigma_{NXS}^2)} - F_{var}
\end{equation}
with
\begin{equation}
     err(\sigma_{NXS}^2) = \sqrt{\left( \sqrt{\frac{2}{N}}\frac{<\sigma_{err}^2>}{<F_\gamma>^2}\right)^2 + \left( \sqrt{\frac{<\sigma_{err}^2>}{N}}\frac{2F_{var}}{<F_\gamma>}\right)^2}\,.
\end{equation}

Considering the whole time epoch from 2022-03-01 to 2022-07-19 (MJD 59639 to MJD 59779), all wavebands show small variability scales. As shown in Figure~\ref{fig:fvar} and reported in Table~\ref{tab:Fvar}, $F_\text{var}$ is always below 0.3 for the full time epoch (closed markers). The high-energy X-rays and VHE $\gamma$-rays display the highest degrees of variability. As reported with data from other multi-instrument campaigns Mets\"ahovi systematically shows a higher $F_{var}$ than OVRO \citep{Mrk501_MAGIC_2008,Mrk501_MAGIC_2013,Mrk501_MAGIC_2012,2021arXiv211011522M}, however, we also find a higher degree of variability of the 230 GHz IRAM data compared with the 86 GHz data. The high variability in the \mbox{37 GHz} is not caused by a flaring event, but rather by a flickering in the radio fluxes, which may perhaps be produced by the somewhat limited sensitivity of Mets\"ahovi, and the relatively low brightness of Mrk\,501 at radio. To check that the higher variability degree in Mets\"ahovi is not caused by a threshold effect, we computed it using different signal quality cuts, namely the signal to noise ratio S/N $>$ 5, 7 and 10. We found that, for all the above-mentioned signal qualities, the computed $F_\text{var}$ values are consistent with the one computed with the full data set (S/N $>$ 4), which is higher than the $F_\text{var}$ obtained for low-frequency radio instruments. Additionally, we investigate the variability for only the \textit{IXPE} time windows shown by the turquoise markers in Figure~\ref{fig:fvar}. Due to a lack of simultaneous data, it is not computed for the $\gamma$-ray data. For the X-rays a higher degree of variability is found than in the full time period, while for the lower wavebands, radio to UV, a very similar behavior between the time epochs is found.

\begin{figure}
\centering
  \resizebox{\hsize}{!}{\includegraphics{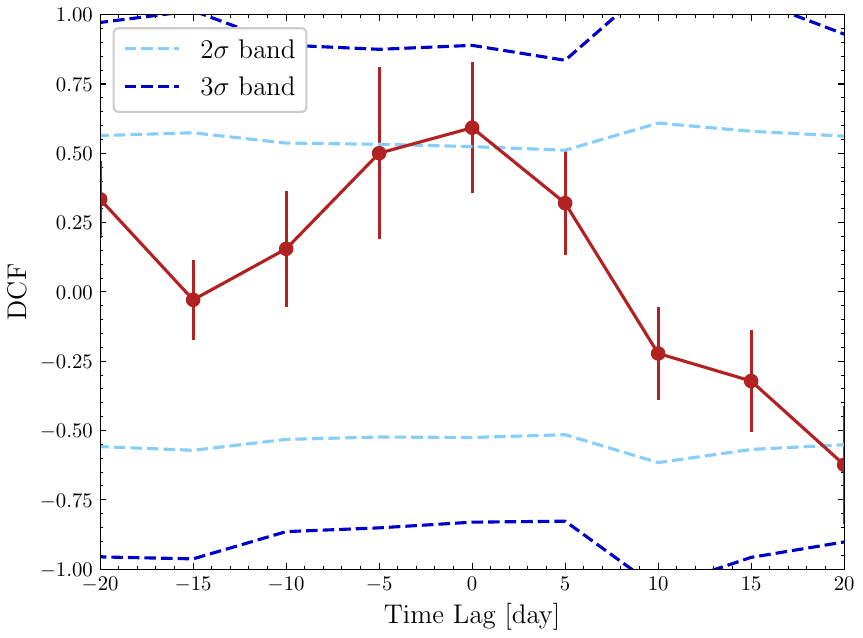}}
  \caption{Correlation between the MAGIC flux above 200~GeV and the \textit{Swift}-XRT flux 0.3--2\,keV. See text for further details.}
  \label{fig:magic_vs_03_2keV}
\end{figure}

\begin{figure}
\centering
  \resizebox{\hsize}{!}{\includegraphics{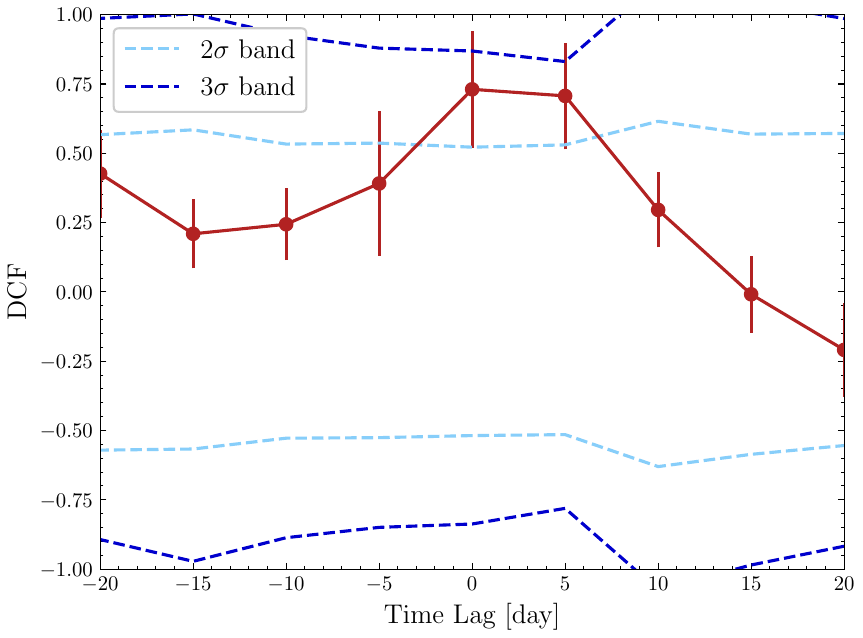}}
  \caption{Correlation between the MAGIC flux above 200~GeV and the \textit{Swift}-XRT flux 2--10\,keV. See text for further details.}
  \label{fig:magic_vs_2_10keV}
\end{figure}

We searched for intra-band correlation and found a hint of positive correlation only between the VHE and X-ray light curves, which also correspond to the energy bands with the strongest variability and the locations of the two SED peaks. The correlation is quantified using the discrete correlation function \citep[DCF;][]{1988ApJ...333..646E}, which we compute between the MAGIC fluxes above 200\,GeV and the  0.3--2\,keV \& 2--10\,keV fluxes from \textit{Swift}-XRT. The DCF is calculated for a series of time lags from -20\,days to +20\,days between the light curves using time lag steps of 5\,days, which is in agreement with the time scale taken from the auto-correlation behavior of Mrk\,501 during low activity in \citet{2023ApJS..266...37A}. The significance of the correlation is estimated based on Monte Carlo simulations in a similar fashion as the one described in \citet{2021A&A...655A..89M}. We simulate $10^4$ pairs of uncorrelated light curves that respect the sampling and binning of the real data. The DCF significance band in each time lag is then obtained from the distribution of DCF values of the simulations. To simulate the light curves, we assume that for each energy band the power spectral density function used as an input for the simulation follows a power-law model with an index of 1.4, which is the same index derived in \citet{2015A&A...573A..50A}. As shown in Figure~\ref{fig:magic_vs_03_2keV} and Figure~\ref{fig:magic_vs_2_10keV} the significance is slightly above $2\sigma$ at zero time lag.


In summary, we find that during the full campaign Mrk\,501 has showed very typical behaviors considering its MWL flux values. It does not just stay around its average state \citep{Abdo_2011}, but also depicts the highest $F_\text{var}$ in the X-rays and VHE as previously found in \citet{Mrk501_MAGIC_2013} and \citet{2023ApJS..266...37A} and shows evidence for the commonly observed correlation between the X-ray and VHE band \citep{Mrk501_MAGIC_2014,Mrk501_MAGIC_2013,Mrk501_MAGIC_2012, 2023ApJS..266...37A}.\par 

As for the long-term evolution in the polarization, we note an increase of the polarization degree on a few day timescale in the R-band around 2022-05-21 (MJD~59720). The R-band polarization increases by a factor $\sim2$ and reaches a level of 12\%, which is above the degree in the X-ray during the \textit{IXPE} epochs. No significant flare is observed simultaneous to this elevated optical polarization state. The lack of simultaneous MWL polarization data prevents us from making more detailed investigations. As for the polarization angle, moderate variability by a few tens of degrees can be seen in the radio and optical bands. A significant offset between the R-band and radio (SMA; 226\,GHz) angle temporarily occurs from 2022-05-31 (MJD~59730) to 2022-06-30 (MJD~59760). The angle deviates by $\sim60^\circ$ around 2022-05-31 (MJD~59730). This behavior contrasts with the one during the \textit{IXPE} epochs where a rough alignment (within $\sim30^\circ$) of the optical/radio/X-ray polarization angle can be noticed.

\section{Modeling of the SEDs during the \textit{IXPE} observing windows}
\label{sec:modelling}

\begin{figure}[h]
\centering
 \includegraphics[width=0.9\columnwidth]{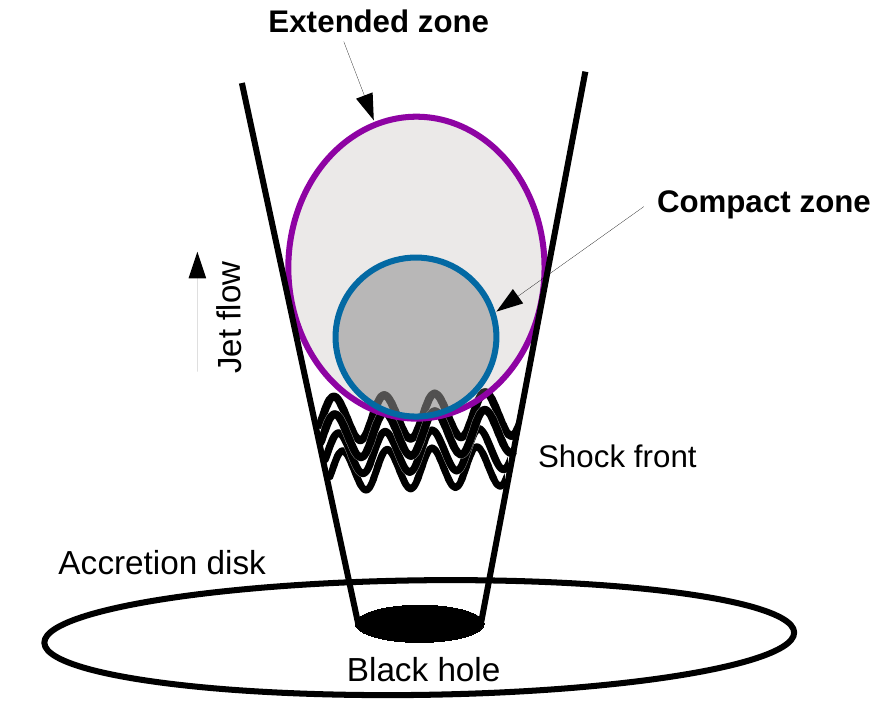}
  \caption{Simplified sketch representing the morphology of the two-zone leptonic model discussed in Section~\ref{sec:modelling}.}
  \label{fig:jet_sketch}
\end{figure}

We exploit the broad-band SEDs shown in Figure~\ref{fig:SED} to perform a modeling of the emission assuming a leptonic scenario. We consider an energy-stratified emission region, as suggested by the optical-to-X-ray polarization properties (see Section~\ref{sec:pol_vs_freq}). The region consists of two overlapping blobs, each of them with different radius $R'$. The two blobs contribute
dominantly to distinct energy regions of the SED. The X-ray and $>100$\,GeV $\gamma$-ray fluxes are dominated by a ``compact zone'', located close to the shock. The second zone, dubbed as ``extended zone'', contributes to the radio, optical/UV and MeV-GeV fluxes. The ``extended zone'' is significantly larger than the ``compact zone'', and expands over a larger volume downstream the shock front. The theoretical framework that we use here is in line with the energy-stratified model of  \citet{2022Natur.611..677L,2016MNRAS.463.3365A, 2016ApJ...833...77I}, where the X-ray emission is produced closer to the shock front than the optical emission. In Figure~\ref{fig:jet_sketch} we show a sketch depicting the jet morphology of our modeling framework. While considering two distinct zones is surely a simplification of the system (one would rather expect a continuum of several regions contributing to the different parts of the SED), a more realistic description of the system would require a much more detailed treatment of particle cooling and diffusion/advection that is beyond the scope of this work. This simplified model already allows one to constrain important properties of the emission states of Mrk\,501 during the three \textit{IXPE} pointings.\par 
In order to limit the degrees of freedom, we make the following assumptions. Parameters referred to the blob frame are marked as primed, while unmarked ones relate to the observer frame.

\begin{itemize}

    \item The Doppler factor $\delta$ is fixed to 11 consistent with the one used in \citet{2023ApJS..266...37A}. For simplicity, the same value is chosen for both regions. 
    
    \item In order to estimate the relative size between the two zones, we proceed as follows: we first assume that the ``compact region'' can be modelled by $N$ turbulent plasma cells of identical magnetic field strength, but with random orientation. In such a configuration, the expected average polarization degree from the zone can be approximated as $P_\text{deg} \approx 75\%/\sqrt{N}$ \citep{2014ApJ...780...87M, 2021Galax...9...37T}. Given that \textit{IXPE}-1 and \textit{IXPE}-2 are characterized by $P_\text{deg, X-ray} \approx 10\%$ in the X-ray band, it implies $N\approx60$ during the latter two epochs. Regarding \textit{IXPE}-3, $P_\text{deg, X-ray}\approx 7\%$, corresponds to $N\approx110$. The size of the ``extended zone'' dominating the optical emission is then assumed to be a factor $l$ greater than the ``compact region'' such that the number of turbulent cells in the ``extended zone'' matches the optical polarization degree according to $P_\text{deg} \approx 75\%/\sqrt{l \, N}$. For \textit{IXPE}-1 and \textit{IXPE}-2, $P_\text{deg, opt.} \approx 5\%$, yielding $l\approx5$. For \textit{IXPE}-3, $P_\text{deg, opt.} \approx 2\%$, which implies $l\approx10$.
    
    Assuming that turbulent cells roughly span equal volumes, we obtain that the radius of the ``extended zone'' should be a factor $l^{1/3}$ larger than the one of the ``compact zone''. For all epochs, we set the radius of the ``compact zone'' to \mbox{$R' = 2.9 \times 10^{16}$\,cm.} This is in agreement with a light crossing time $R'/(c\delta)\approx1$\,day, which is the variability timescale observed in the X-ray band.

    \item The electron distribution inside the ``compact zone'' is modelled with a broken power-law function normalized such that the electron energy density is given by $U'_e$. The distribution is defined between a minimum and maximum Lorentz factor of $\gamma'_{min}$ and $\gamma'_{max}$. The break is located at $\gamma'_{break}$. The slopes below and above the break are given by $n_1$ and $n_2$, respectively. $\gamma'_{min}$ is set to a value such that the ``compact zone'' remains sub-dominant in the optical/UV. With the choice of $\delta$ and $R'$ mentioned above, we derive $\gamma'_{min} \sim 10^4$. Within an ion-electron plasma, mildly relativistic shocks, in which the shock front is moving at a Lorentz factor $\gamma_{sh} \sim 1-3$ relative to the unshocked plasma, are expected to generate somewhat lower $\gamma'_{min}$, in the order of a few $10^3$ \citep{2021A&A...654A..96Z}. Values of $\gamma'_{min} \sim 10^4$ may be reached for fully relativistic shocks with $\gamma_{sh}$ of a few tens. In order to constrain $\gamma'_{max}$, we equate the acceleration and cooling timescales. The acceleration timescale in shock acceleration is estimated with:
    \begin{equation}
    t'_{acc} = \frac{20 \lambda(\gamma') c }{3 u^2_s} 
    \end{equation}

    where $\lambda(\gamma') = (\xi \gamma' m_e c^2)/(eB')$ is the mean free path of electrons, parameterized as a fraction $\xi$ of the Larmor radius. $\xi$ is an acceleration parameter, which we fix to $10^4$. The latter value is in agreement with previous estimates from \citet{2002MNRAS.337..609Z} based on spectral hysteresis observations of a similar HSP, Mrk~421. $B'$ is the magnetic field strength and $e$ the electron charge, $m_e$ the electron mass. $u_s$ is the speed of the shock, that we assume to be relativistic, $u_s \sim c$. The synchrotron cooling timescale is:
    \begin{equation}
    t'_{cool, synch} = \frac{6 \pi m_e c}{\sigma_T B'^2 \gamma'} 
\end{equation}
where $\sigma_T$ is the Thomson cross section.

    \item The electron distribution inside the ``extended zone'' is modelled with a simple power-law function because the available data prevent a precise determination of breaks or other distributions with more degrees of freedom. The slope of the distribution is assumed to be $p=2.2$. The minimum and maximum Lorentz factor of the distribution, $\gamma'_{min}$ \& $\gamma'_{max}$ are constrained by the radio-to-UV data. The distribution is normalized to an energy density of $U'_e$. We keep $\gamma'_{max}$ within a factor 2 from the expected cooling break that is given by equating the cooling timescale due to synchrotron and inverse-Compton processes with the escape timescale $R'/c$.
    
\end{itemize}

We start by first fitting the X-ray and $\gamma$-ray data to determine the remaining free parameters of the ``compact zone'' (i.e., magnetic field $B'$, $U'_e$, $n_1$, $n_2$). In a second step, the ``extended zone'' is added in order to get a good description of the radio-to-UV and \textit{Fermi}-LAT data. Between the \textit{IXPE-1} and \textit{IXPE}-2 epochs, the data are satisfactorily described with the same parameters for the ``extended zone'', which is not surprising given the lower variability in the radio, optical/UV and \textit{Fermi}-LAT data (see Figure~\ref{fig:MWL_LC} and Figure~\ref{fig:fvar}), and the relatively short time period between \textit{IXPE}-1 and \textit{IXPE}-2 of a few weeks, while \textit{IXPE}-3 took place several month later. \par

\begin{figure}
        \centering
        \begin{subfigure}[b]{0.95\columnwidth}
            \centering
            \includegraphics[width=1.03\linewidth]{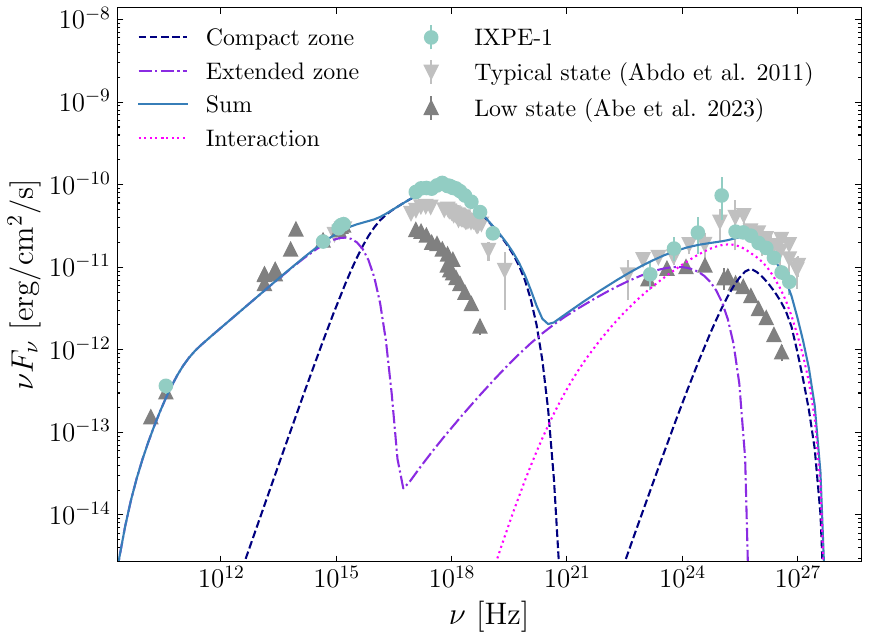}
        \end{subfigure}
        \begin{subfigure}[b]{0.95\columnwidth}  
            \centering 
            \includegraphics[width=1.03\linewidth]{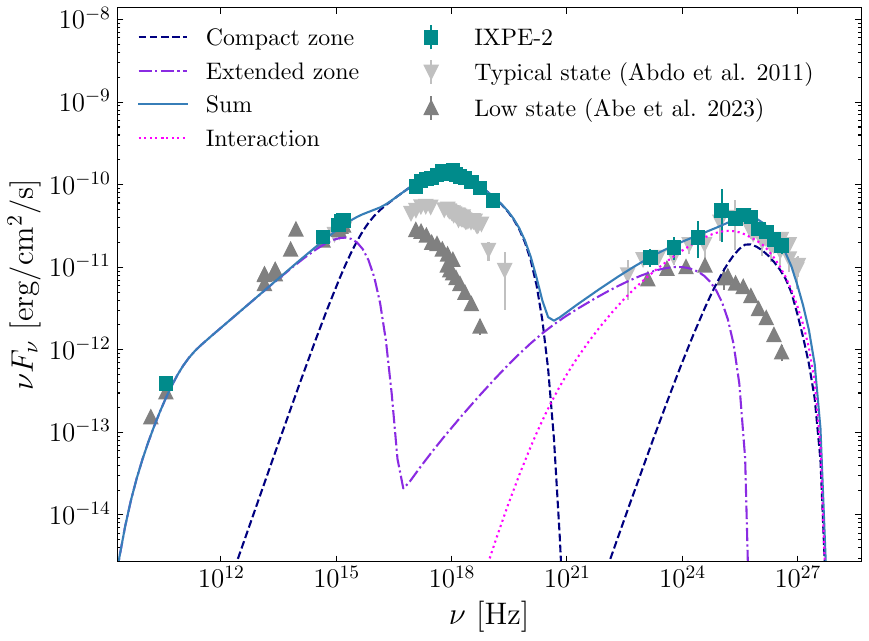}
        \end{subfigure}
        \begin{subfigure}[b]{0.95\columnwidth}
            \centering
            \includegraphics[width=1.03\linewidth]{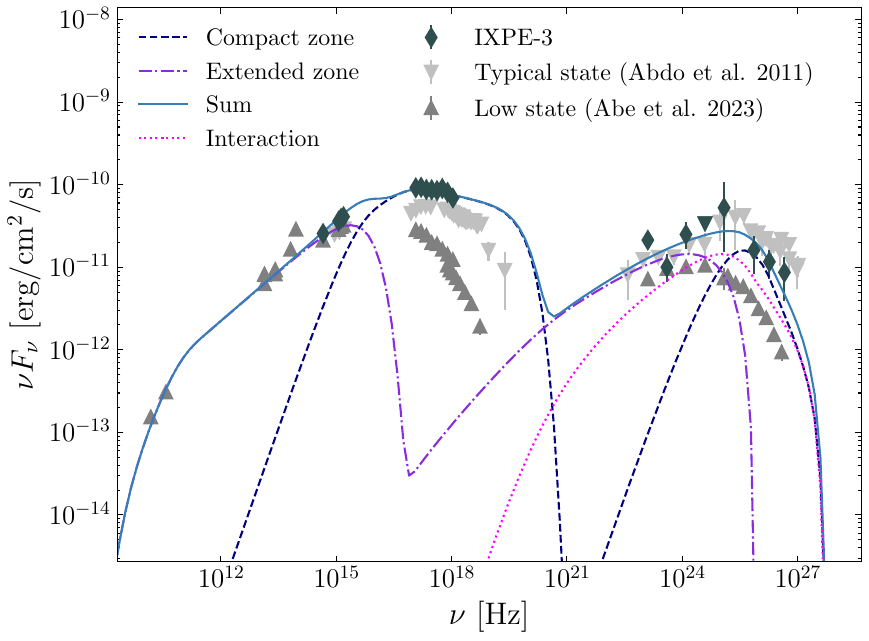}
        \end{subfigure}
        \caption{Modelling of the three \textit{IXPE} epochs (Top: \textit{IXPE}-1; Middle: \textit{IXPE}-2; Bottom: \textit{IXPE}-3). The dashed blue curve is the emission originating from the ``compact zone'', located nearby the shock front. The dashed-dotted violet curve is the contribution from the ``extended zone'', which spans a larger volume downstream the shock. The emission resulting from the interaction between the two zones is plotted in pink dotted curve. The sum of all components is given by the solid blue line. The light grey and dark grey markers show the average state \citep[from][]{Abdo_2011} and low state \citep[from][]{2023ApJS..266...37A}, respectively. The parameters of the model are listed in Table~\ref{tab:ssc_parameters}. } 
\label{fig:IXPE_twozones}
\end{figure}

The adopted model is shown in Figure~\ref{fig:IXPE_twozones} for each epoch.
The dashed blue lines and dashed-dotted violet lines are the contributions from the ``compact zone'' and ``extended zone'', respectively. The solid light blue line is the sum of the two regions. We also compute the interaction between the two zones (dotted pink line), which results from electron inverse-Compton scattering off the photon field that the two regions feed to each others. The derived parameter values are shown in Table~\ref{tab:ssc_parameters}. Overall, the model describes well all three broad-band SEDs. A more detailed interpretation of the results is given in Section~\ref{sec:discussion}.\par

\begin{table*}
\caption{\label{tab:ssc_parameters}Parameters of the two-components of the leptonic model obtained for the three \textit{IXPE} epochs.}
\centering
\begin{tabular}{l c c c c @{\hskip 0.3in} c c c}    
\hline\hline
\multirow{2}{*}{Parameters}   & \multicolumn{3}{c}{``compact zone''} &  &  & ``extended zone'' & \\\cline{2-4}\cline{6-8} 
  & \textit{IXPE}-1 & \textit{IXPE}-2 & \textit{IXPE}-3 & & \textit{IXPE}-1 & \textit{IXPE}-2 & \textit{IXPE}-3\\  
\hline
\hline
$B'$ [$10^{-2}$G]  & 5.0 & 5.0 & 6.8 & & 3.5 & 3.5 & 3.2 \\
$R'$ [$10^{16}$cm] & 2.9 & 2.9 & 2.9 & & 5.0 & 5.0 & 6.3 \\
$\delta$ & 11 & 11 & 11 & & 11 & 11 & 11  \\
$U'_e$ [$10^{-3}$\,erg cm$^{-3}$] & 0.8 & 1.2 & 0.8 & & 2.8 & 2.8 & 2.0\\
$n_1$ & 2.37 & 2.25 & 2.20 & & 2.2 & 2.2 & 2.2\\
$n_2$ & 4.00 & 3.67 & 3.20 & & (...) & (...) & (...) \\
$\gamma'_{min}$ & $5\times10^{4}$ & $4\times10^{4}$ & $3\times10^{4}$ & & $2\times10^{2}$ & $2\times10^{2}$ & $2\times10^{2}$ \\
$\gamma'_{br}$ & $6.0\times10^{5}$ & $6.0\times10^{5}$ & $1.6\times10^{5}$ & & (...) & (...) & (...) \\
$\gamma'_{max}$ & $5.5\times10^{6}$ & $5.5\times10^{6}$ & $4.8\times10^{6}$ & & $5.7\times10^{4}$ & $5.7\times10^{4}$ & $7.2\times10^{4}$ \\
\hline 
$U'_e/U'_B$ & 8 & 12 & 4 & & 57 & 57 & 50 \\
\hline
\hline
\end{tabular}
\tablefoot{ We refer the reader to Section~\ref{sec:modelling} for a detailed description of each parameter.}
\end{table*}

\section{Summary and discussion} \label{sec:discussion}

In this work, we present the multi-wavelength picture around the first X-ray polarization measurements of Mrk\,501. For the first time, simultaneous information in the VHE regime is published and compared to the X-ray measurements allowing the first broad-band SEDs up to the VHE band to be constructed covering the three \textit{IXPE} time windows (\textit{IXPE}-1 and \textit{IXPE}-2 from 2022-03-08 to 2022-03-10 and 2022-03-27 to 2022-03-29 (MJD 59646 to 59648 and MJD 59665 to 59667) and \textit{IXPE}-3 from 2022-07-09 to 2022-07-12 (MJD 59769 to 59772). 

Both the VHE and X-ray regimes showed the highest flux level during \textit{IXPE}-2 ($\sim$ factor of 2 higher than during the other \textit{IXPE} windows). Comparing the spectral properties, a spectral hardening with increasing flux levels is observed in the \mbox{X-rays} between the different pointings, which is supported by the ``harder when brighter'' trend observed over the full campaign (see Figure~\ref{fig:xrt_hardness}). The opposite behavior can be seen in the high-energy $\gamma$-ray data where the spectral index is softening with higher flux levels. No spectral change in the VHE is seen between the three pointings. Interestingly, the power-law indices in the X-rays are rather hard during \textit{IXPE}-1 and \textit{IXPE}-2 ($\lesssim$2), as is typically observed for elevated flux states, while the VHE spectral hardness remains close to the 
average Mrk~501 activity \citep[power-law index $\sim$2.5;][]{Abdo_2011}. 

Noteworthy is not just the change in spectral index, but also the broad-band characteristics of the different SEDs (see Table~\ref{tab:peak_freq}), which revealed an EHSP behavior of Mrk\,501 in the \textit{IXPE}-1 and \textit{IXPE}-2 states with a synchrotron peak frequency well beyond 1\,keV. During \textit{IXPE}-3 the blazar has gone back to its more typical HSP behavior shifting the synchrotron peak by an order of magnitude towards lower energies. The shift in synchrotron peak frequencies can explain the positive change in flux observed in the UV band of the MWL LC around April (see Figure~\ref{fig:MWL_LC}). The accompanying shift of the high-energy peak can explain the different flux behavior of the high-energy $\gamma$-rays, which increases steadily between the three states compared with the rise and then fall of the fluxes in the VHE and X-rays.
During all three states, we find a low Compton dominance compared to the typical state of Mrk\,501 \citep{Abdo_2011} by a factor of 2 (see Table~\ref{tab:peak_freq}). Additionally, we show the same parameters obtained from the leptonic modeling in Table~\ref{tab:peak_freq_model} for comparison. While the parameters for \textit{IXPE}-1 and \textit{IXPE}-2 match very well, both peak frequencies shift towards lower energies for \textit{IXPE}-3. However, when considering the increased uncertainties of the parameters for the \textit{IXPE}-3 state, a clear shift can only be claimed for the synchrotron peak. This shows the poorer coverage of this emission state in our data, and highlights that especially the perceived drop in CD for \textit{IXPE}-3 as observed in Table~\ref{tab:peak_freq} should be considered with caution due to the dependence on the peak frequency.

\begin{table}[h]
\caption{Peak frequencies, $\nu_\text{s}$ and $\nu_\text{IC}$, and Compton dominance (CD) for the different SEDs shown in Fig.~\ref{fig:SED} extracted from the SSC description described in Section~\ref{sec:modelling}.}
\centering
\begin{tabular}{ l | c c c}  
\multirow{2}{*}{States} & $\nu_\text{s}$ & $\nu_\text{IC}$  & CD\\
 &  [Hz] & [Hz] & \\
\hline \hline
\textit{IXPE}-1$_\text{SSC}$ &  $6.3 \times10^{17} $ & $4.4 \times10^{25}$ & 0.26\\
\textit{IXPE}-2$_\text{SSC}$ &  $9.0 \times10^{17}$ & $4.4 \times10^{25}$ & 0.29 \\
\textit{IXPE}-3$_\text{SSC}$ &  $1.6  \times10^{17}$ & $1.5\times10^{25}$ & 0.31 \\
\end{tabular}
\label{tab:peak_freq_model}
\end{table}

Mrk\,501 has previously shown EHSP behavior during large flaring activity in 1997 \citep{2001ApJ...554..725T} and 2013 \citep{Mrk501_MAGIC_2013}. This EHSP behavior during large flaring activity has also been seen in other HSP blazars \citep{2020MNRAS.496.3912M,2020A&A...638A..14M}. Additionally, similarly to the HSP 1ES~2344+514 \citep{2023arXiv231003922A}, Mrk\,501 has shown EHSP behavior during non-flaring activity, as reported in \citet{Mrk501_MAGIC_2012}, that used the extensive multi-instrument campaign in 2012. Compared to the data from 2022 that is reported in this manuscript, the above-mentioned studies from past campaigns showed a far higher variability in both VHE and X-rays. While similar average flux levels are found in the X-rays in 2012 compared to this work, the average VHE flux levels were close to the typical flux levels of Mrk\,501 during the EHSP behavior. However, both bands depict larger flux changes in 2012 than in the 2022 campaign, together with higher degrees of variability, reaching up to $F_{var}\sim 1$ in the VHE. The main difference in behavior during 2022 compared to that reported in 2012/2013 is that only the X-rays show a hard spectrum while in 2012/2013 both X-rays and VHE revealed atypically hard SEDs.

In addition to shift in peak frequencies, the polarization degree drops for the \textit{IXPE}-3 pointing, both in the X-rays  and optical regime. However, the ratio between the optical and X-ray band polarization stays rather constant during all three pointings (see Figure~\ref{fig:pol_deg_evol}). In fact, the X-ray polarization is systematically a factor $\sim2-3$ higher than in the optical band for each \textit{IXPE} epoch. No obvious trend between the changes in flux in the corresponding regimes and the change in polarization is seen. Therefore, the spectral changes and broad-band SEDs are more probable to explain the differences in polarization degree between the different states as will be discussed further below using our theoretical modeling.\par 

The polarization angle during the \textit{IXPE} epochs shows a rough alignment between the radio, optical and X-ray bands (within $\sim30^\circ$). The values are close to the average jet angle of $119\pm12^\circ$ determined by \citet{Weaver_2022} using Very Long Baseline Array imaging at 43\,GHz (see Figure~\ref{fig:MWL_LC}, bottom panel). Nonetheless, in the middle of the campaign, both the radio (SMA; 226\,GHz) and optical (R-band) polarization angles depict some variability and deviate from the jet angle in an incoherent manner. Indeed, around 2022-05-31 (MJD~59730), the offset between the radio and optical bands reaches $\sim60^\circ$. This behaviour strongly points towards a separation (at least partially) between the optical and radio regions. The deviation of the polarization angle from the jet axis may arise if the emitting regions are located in a portion of the jet where bending on short scale occurs \citep{2005MNRAS.360..869L,2018A&A...619A..88M,2021Galax...9...37T}. Furthermore, it points towards multiple emission regions being responsible for the long-term behavior of Mrk\,501 with a stable part of the magnetic field regulated by shocks, but at times dominated by more variable/turbulent contributions as also proposed in \citet{2023ApJS..266...37A}.

Over the full time period, the blazar showed a rather stable behavior with typical \citep{Abdo_2011} flux levels in the VHE ($\sim$20-50\% C.U.). In contrast, the XRT flux values are constantly above $10^{-10}$\,erg\,cm$^{-2}$\,s$^{-1}$. 
Based on the long-term light curve presented in \citet{2023ApJS..266...37A}, this indicates an elevated state with respect to the average X-ray activity \citep{Abdo_2011}. It confirms that the atypically low Compton dominance during the \textit{IXPE} windows is a feature present throughout the entire campaign presented in this work. For all bands, the fractional variability stays below $F_{var} =0.3$ when considering the full time epoch with slightly higher ones in the X-rays for only the \textit{IXPE} time windows (see Figure~\ref{fig:fvar}). Even though different behaviors are reported in the X-rays and VHE, we identify evidence for a correlation between the two bands without a time lag (see Figures~\ref{fig:magic_vs_03_2keV} and \ref{fig:magic_vs_2_10keV}). The correlations are found with significances between $2\sigma$ to $3\sigma$, limited by the short time epoch considered, the low flux levels, the very low variability, and the somewhat limited sensitivity in the VHE $\gamma$-ray band in comparison with the higher sensitivity to measure flux changes in the X-ray band.\par

To further investigate the emission states behind the first \textit{IXPE} pointings, we modelled the broad-band SEDs with a two-zone leptonic model (see Section~\ref{sec:modelling}). We assume a ``compact'' region, which is nearby the shock front and populated by freshly accelerated electrons, responsible for the X-ray and VHE \mbox{$\gamma$-ray} band. The ``compact region'' is embedded into an ``extended region'' which expands downstream the shock. The ``extended region'' dominates the emission in the radio, optical, UV and \textit{Fermi}-LAT bands. Such a morphology of the emitting region is motivated by the energy dependency of the polarization degree between the radio and the X-rays \citep{2022Natur.611..677L}, which points towards an energy stratified jet. In addition, previous long-term correlation studies \citep[e.g.][]{2023ApJS..266...37A} points towards the same emitting region dominating the X-ray and \mbox{$\gamma$-ray} bands. While the ``extended'' zone is assumed to be constant between the three pointing, the ``compact'' one varies over time to describe the observed spectral changes.\par 

A good description of the observations from radio to VHE is achieved for each epoch. The X-ray data during \textit{IXPE}-1 and \textit{IXPE}-2 constrain well both the rising and falling edges of the synchrotron SED close to the peak energy. This provides in turn strong constraints on the spectral shape of the electron distribution within the region nearby the shock front. Below the break, which either results from cooling or acceleration effects, the derived index of the electron distribution is $n_1=2.25$ for the ``compact zone''. This is well in agreement with expectation from diffusive acceleration in relativistic shocks \citep{2000ApJ...542..235K}. Regarding the \textit{IXPE}-3 epoch, similar $n_1$ are obtained although in those cases $n_1$ is less constrained given the shift of the synchrotron peak towards lower energies.\par

The emission resulting from the interaction between two zones brings a significant contribution to the \textit{Fermi}-LAT band. Consequently, the X-rays should not only correlate with the VHE fluxes (as reported several times in the past for Mrk~501), but also with the one from \textit{Fermi}-LAT. Using 12 years of data, \citet{2023ApJS..266...37A} detected a positive correlation between the \textit{Swift}-XRT and \textit{Fermi}-LAT emission, which is thus in good agreement with our model.\par

Between the different epochs, most of the parameters of the ``compact zone'' are quite similar. A main difference is the drop of $\gamma'_{br}$ by a factor $\approx4$ during \textit{IXPE}-3 with respect to \textit{IXPE}-1 and \textit{IXPE}-2 in order to explain the shift towards lower energies of the synchrotron component. The obtained $\gamma'_{break}$ for \textit{IXPE}-1 and \textit{IXPE}-2 are a factor of 4 higher than the one expected for a self-consistent equilibrium between cooling due to synchrotron and inverse-Compton processes, injection and escape \citep{Tavecchio_Constraints}. For \textit{IXPE}-3 the difference relaxes to a factor of 2. Additionally, the magnetic field $B'$ increases by $\approx 40\%$ during \textit{IXPE}-3, from 0.050\,G to 0.068\,G. \par

The drop in the optical/X-ray polarization degree during the \textit{IXPE}-3 epoch compared to the earlier \textit{IXPE} epochs suggests a general evolution of the shock properties. The increase of $B'$ in the ``compact zone'' for \textit{IXPE}-3 points towards an evolution of the magnetization. Within a multi-cell scenario, the shift in peak frequencies can be attributed to an increase of synchrotron cooling due to stronger magnetic field strength. At the same time, the drop in polarization implies that the magnetic field becomes less ordered and that the emitting region is composed by a larger number of turbulent plasma cells. It has been shown before that turbulences created by relativistic shock propagation could lead to an amplification in the magnetic field \citep{2014MNRAS.439.3490M}. Furthermore, as discussed in \citet{2000ApJ...542..235K} and \citet{2011MNRAS.414.2017K}, a higher magnetization can lead to a decrease of the shock compression ratio $\kappa$ (defined as the ratio between the upstream and downstream plasma velocity in the shock reference frame). Since the polarization degree is positively correlated with $\kappa$, it can be expected that an increase of the magnetization caused by a more turbulent plasma results in an overall decrease of the polarization degree. Such a scenario may thus explain why \textit{IXPE-3} is characterized by a larger number of turbulent cells compared to \textit{IXPE-1} and \textit{IXPE-2} as pointed out in Sect.~\ref{sec:modelling}, where we derive $\sim60$ cells for \textit{IXPE-1} and \textit{IXPE-2} and $\sim110$ cells for \textit{IXPE-3} in the ``compact zone''.\par  

For simplicity, the radius of the ``compact zone'' is the same for all epochs in our model and is set to the highest value allowed by the daily timescale variability observed in the X-rays. However, one may argue that a higher number of turbulent cells during \textit{IXPE-3} actually suggests a larger emitting zone with respect to the other epochs. If one assume a stable magnetic field of $5\times 10^{-2}$\,G in the ``compact zone'' throughout the observing campaign (which is the value derived for \textit{IXPE}-1 and \textit{IXPE}-2), we find that the radius of the ``compact zone'' should increase by $\sim$20\% with respect to \textit{IXPE-1} and \textit{IXPE-2} to properly describe the SED of \textit{IXPE}-3. Such a relative radius increase leads to $1.7$ times more turbulent cells in the emitting region, under the assumption that turbulent cells have a roughly constant volume over time. This is in good agreement with the fact that the \mbox{\textit{IXPE-3}} ``compact zone'' should be composed by $\sim$twice the number of turbulent cells compared to \textit{IXPE-1} and \textit{IXPE-2} (see Sect.~\ref{sec:modelling}). In summary, based on our modeling, the decrease of the polarization degree throughout the \textit{IXPE} epochs may be qualitatively explained by a change of magnetization and/or a change of the emitting region size.\par

In conclusion, with this 2022 data set we could, for the first time, combine broad-band MWL data up to the VHE together with polarization measurements up to the X-rays, which allowed us to better constrain the emission and acceleration behind the observed radiation. This shows the crucial role of further MWL monitoring in combination with coverage of the X-ray polarization of different emission states of Mrk\,501 to extend our knowledge on the mechanisms behind the blazar's emission. Further insights could be gained by polarization measurements at even higher energies covering the high-energy peak of the SED. In this context, e-ASTROGAM \citep[][]{eAstrogram}, COSI \citep{COSI} or AMEGO \citep{amego}, which are being constructed or considered for construction in the next years, could add even more vital insights.

\section*{Author contribution}

A. Arbet Engels: project management, coordination of MWL data analysis,  \textit{NuSTAR} analysis, correlation analysis, theoretical modeling and interpretation, paper drafting; L. Linhoff: MAGIC analysis cross-check; I. Liodakis: organization of \textit{IXPE} and MWL observations, coordination of \textit{IXPE}  and MWL data; L. Heckmann: project management, coordination of MWL data analysis, MAGIC and Fermi data analysis, variability analysis, theoretical interpretation, paper drafting; D. Paneque: organization of the MWL observations, theoretical interpretation, paper drafting; The rest of the authors have contributed in one or several of the following ways: design, construction, maintenance and operation of the instrument(s) used to acquire the data; preparation and/or evaluation of the observation proposals; data acquisition, processing, calibration and/or reduction; production of analysis tools and/or related Monte Carlo simulations; overall discussions about the contents of the draft, as well as related refinements in the descriptions.

\begin{acknowledgements}
The MAGIC collaboration would like to thank the Instituto de Astrof\'{\i}sica de Canarias for the excellent working conditions at the Observatorio del Roque de los Muchachos in La Palma. The financial support of the German BMBF, MPG and HGF; the Italian INFN and INAF; the Swiss National Fund SNF; the grants PID2019-104114RB-C31, PID2019-104114RB-C32, PID2019-104114RB-C33, PID2019-105510GB-C31, PID2019-107847RB-C41, PID2019-107847RB-C42, PID2019-107847RB-C44, PID2019-107988GB-C22 funded by the Spanish MCIN/AEI/ 10.13039/501100011033; the Indian Department of Atomic Energy; the Japanese ICRR, the University of Tokyo, JSPS, and MEXT; the Bulgarian Ministry of Education and Science, National RI Roadmap Project DO1-400/18.12.2020 and the Academy of Finland grant nr. 320045 is gratefully acknowledged. This work was also been supported by Centros de Excelencia ``Severo Ochoa'' y Unidades ``Mar\'{\i}a de Maeztu'' program of the Spanish MCIN/AEI/ 10.13039/501100011033 (SEV-2016-0588, CEX2019-000920-S, CEX2019-000918-M, CEX2021-001131-S, MDM-2015-0509-18-2) and by the CERCA institution of the Generalitat de Catalunya; by the Croatian Science Foundation (HrZZ) Project IP-2016-06-9782 and the University of Rijeka Project uniri-prirod-18-48; by the Deutsche Forschungsgemeinschaft (SFB1491 and SFB876); the Polish Ministry Of Education and Science grant No. 2021/WK/08; and by the Brazilian MCTIC, CNPq and FAPERJ.\\

The Imaging X-ray Polarimetry Explorer ({\it IXPE}) is a joint US and Italian mission. 
The US contribution is supported by the National Aeronautics and Space Administration (NASA) and led and managed by its Marshall Space Flight Center (MSFC), with industry partner Ball Aerospace (contract NNM15AA18C). 
The Italian contribution is supported by the Italian Space Agency (Agenzia Spaziale Italiana, ASI) through contract ASI-OHBI-2017-12-I.0, agreements ASI-INAF-2017-12-H0 and ASI-INFN-2017.13-H0, and its Space Science Data Center (SSDC), and by the Istituto Nazionale di Astrofisica (INAF) and the Istituto Nazionale di Fisica Nucleare (INFN) in Italy.
This research used data products provided by the {\it IXPE} Team (MSFC, SSDC, INAF, and INFN) and distributed with additional software tools by the High-Energy Astrophysics Science Archive Research Center (HEASARC), at NASA Goddard Space Flight Center (GSFC). 

The \textit{Fermi} LAT Collaboration acknowledges generous ongoing support
from a number of agencies and institutes that have supported both the
development and the operation of the LAT as well as scientific data analysis.
These include the National Aeronautics and Space Administration and the
Department of Energy in the United States, the Commissariat \`a l'Energie Atomique
and the Centre National de la Recherche Scientifique / Institut National de Physique
Nucl\'eaire et de Physique des Particules in France, the Agenzia Spaziale Italiana
and the Istituto Nazionale di Fisica Nucleare in Italy, the Ministry of Education,
Culture, Sports, Science and Technology (MEXT), High Energy Accelerator Research
Organization (KEK) and Japan Aerospace Exploration Agency (JAXA) in Japan, and
the K.~A.~Wallenberg Foundation, the Swedish Research Council and the
Swedish National Space Board in Sweden.
Additional support for science analysis during the operations phase is gratefully 
acknowledged from the Istituto Nazionale di Astrofisica in Italy and the Centre 
National d'\'Etudes Spatiales in France. This work performed in part under DOE 
Contract DE-AC02-76SF00515.

The corresponding authors of this manuscript, namely Axel Arbet-Engels, Lea Heckmann and David Paneque, acknowledge support from the Deutsche Forschungs gemeinschaft (DFG, German Research Foundation) under Germany’s Excellence Strategy – EXC-2094 – 390783311.

The IAA-CSIC group acknowledges financial support from the grant CEX2021-001131-S funded by MCIN/AEI/10.13039/501100011033 to the Instituto de Astrof\'isica de Andaluc\'ia-CSIC and through grant PID2019-107847RB-C44. The POLAMI observations were carried out at the IRAM 30m Telescope. 
IRAM is supported by INSU/CNRS (France), MPG (Germany), and IGN (Spain). Some of the data are based on observations collected at the Observatorio de Sierra Nevada, owned and operated by the Instituto de Astrof\'{i}sica de Andaluc\'{i}a (IAA-CSIC). Further data are based on observations collected at the Centro Astron\'{o}mico Hispano en Andalucía (CAHA), operated jointly by Junta de Andaluc\'{i}a and Consejo Superior de Investigaciones Cient\'{i}ficas (IAA-CSIC). Some of the data reported here are based on observations made with the Nordic Optical Telescope, owned in collaboration with the University of Turku and Aarhus University, and operated jointly by Aarhus University, the University of Turku, and the University of Oslo, representing Denmark, Finland, and Norway, the University of Iceland and Stockholm University at the Observatorio del Roque de los Muchachos, La Palma, Spain, of the Instituto de Astrofisica de Canarias. 
E. L. was supported by Academy of Finland projects 317636 and 320045. 
The data presented here were obtained [in part] with ALFOSC, which is provided by the Instituto de Astrofisica de Andalucia (IAA) under a joint agreement with the University of Copenhagen and NOT. We acknowledge funding to support our NOT observations from the Finnish Centre for Astronomy with ESO (FINCA), University of Turku, Finland (Academy of Finland grant nr 306531). The research at Boston University was supported in part by National Science Foundation grant AST-2108622, NASA Fermi Guest Investigator grants 80NSSC23K1507 and 80NSSC22K1571, and NASA Swift Guest Investigator grant 80NSSC22K0537. T. H. was supported by the Academy of Finland projects 317383, 320085, 322535, and 345899.
This research has made use of data from the RoboPol program, a collaboration between Caltech, the University of Crete, IA-FORTH, IUCAA, the MPIfR, and the Nicolaus Copernicus University, which was conducted at Skinakas Observatory in Crete, Greece.
D.B., S.K., R.S., N.M., acknowledge support from the European Research Council (ERC) under the European Unions Horizon 2020 research and innovation program under grant agreement No.~771282. 
C.C. acknowledges support from the European Research Council (ERC) under the HORIZON ERC Grants 2021 program under grant agreement No. 101040021. 
We acknowledge the use of public data from the Swift data archive. 
Based on observations obtained with XMM-Newton, an ESA science mission with instruments and contributions directly funded by ESA Member States and NASA. 
This research has made use of data from the OVRO 40-m monitoring program (Richards et al. (2011)), supported by private funding from the California Institute of Technology and the Max Planck Institute for Radio Astronomy, and by NASA grants NNX08AW31G, NNX11A043G, and NNX14AQ89G and NSF grants AST-0808050 and AST- 1109911. This publication makes use of data obtained at Mets\"ahovi Radio Observatory, operated by Aalto University in Finland.  AJCT EFG HYD acknowledge support from Spanish MICINN project PID2020-118491GB-I00. This work was supported by JST, the establishment of university fellowships towards the creation of science and technology innovation, Grant Number JPMJFS2129. This work was supported by Japan Society for the Promotion of Science (JSPS) KAKENHI Grant Numbers JP21H01137. This work was also partially supported by the Optical and Near-Infrared Astronomy Inter-University Cooperation Program from the Ministry of Education, Culture, Sports, Science and Technology (MEXT) of Japan. We are grateful to the observation and operating members of the Kanata Telescope. 

The Submillimeter Array is a joint project between the Smithsonian Astrophysical Observatory and the Academia Sinica Institute of Astronomy and Astrophysics and is funded by the Smithsonian Institution and the Academia Sinica. Maunakea, the location of the SMA, is a culturally important site for the indigenous Hawaiian people; we are privileged to study the cosmos from its summit.

This work was supported by NSF grant AST-2109127.

The Joan Or\'o Telescope (TJO) of the Montsec Observatory (OdM) is owned by the Catalan Government and operated by the Institute for Space Studies of Catalonia (IEEC).
H.K. acknowledge NASA support through the grants NNX16AC42G, 80NSSC20K0329, 80NSSC20K0540, NAS8- 03060, 80NSSC21K1817, 80NSSC22K1291, and 80NSSC22K1883.
\end{acknowledgements}

\bibliographystyle{aa}
\bibliography{bibliograhy.bib}

\begin{appendix}
\section{Additional information - Long-term multi-wavelength evolution}
Table~\ref{tab:Fvar} summarizes the values depicted in Fig.~\ref{fig:fvar}.
\begin{table}[h]
\centering
\begin{tabular}{l | c  c }   
Instruments & Full time epoch  & \textit{IXPE} time epochs \\
\hline  \hline
OVRO(15GHz)&0.011$\pm$0.004& (...)\\
Mets\"ahovi (37GHz)&0.131$\pm$0.019&  (...)\\
IRAM (86GHz)&0.028$ \pm $0.017& (...)\\
SMA (226GHz)&0.090$\pm$0.046& (...)\\
IRAM (230GHz)&0.176$\pm$0.019& 0.142$\pm$0.026\\
Optical (R-band)&0.050$\pm$0.010& (...)\\
UVOT (W1)&0.088$\pm$0.005&0.078 $\pm$0.012\\
UVOT (M2)&0.093$\pm$0.005&0.074$\pm$ 0.012\\
UVOT (W2)&0.099$\pm$0.005&0.083$\pm$  0.012\\
XRT (0.3-2keV)&0.136$\pm$0.002&0.187 $\pm$0.005\\
XRT (2-10keV)&0.243$\pm$0.005&0.385$\pm$ 0.011\\
NuSTAR (3-7keV)&0.178$\pm$0.002& (...)\\
NuSTAR (7-30keV)&0.259$\pm$0.003& (...)\\
IXPE (2-8keV)&0.504$\pm$0.017& (...)\\
LAT (0.3-300GeV)&0.145$\pm$0.099& (...)\\
MAGIC (0.2-1TeV)&0.222$\pm$0.029& (...)\\
MAGIC (>1TeV)&0.233$\pm$0.064& (...)\\
\end{tabular}
\caption{Fractional variability $F_{var}$ values displayed in Fig.~\ref{fig:fvar}.}
\label{tab:Fvar}
\end{table}
\end{appendix}
\end{document}